\documentclass[universe,article,accept,moreauthors,pdftex,10pt,a4paper]{mdpi} 

\usepackage{amssymb}

\usepackage{natbib}
\def\beq{\begin{equation}}
\def\eeq{\end{equation}}
\def\bea{\begin{eqnarray}}
\def\eea{\end{eqnarray}}

\def\RS{\Sigma}
\providecommand{\dif}{\mathrm{d}} % upright differential
\def\d{\dif}

\newcommand{\cred}[1]{\textcolor{black}{#1}}
\newcommand{\cblue}[1]{\textcolor{black}{#1}}
\newcommand{\cgreen}[1]{\textcolor{black}{#1}}

\newcommand{\ce}{{\cal{E}}} 
\newcommand{\cl}{{\cal{L}}} 
\newcommand{\cb}{{\cal{B}}}
\newcommand{\BB}{{\cal{B}}}

\newcommand{\apj}{Astrophys. J.}
\newcommand{\apjl}{Astrophys. J. Lett.}
\newcommand{\mnras}{Mon. Not. R. Astron. Soc.}
\newcommand{\prd}{Phys. Rev. D}
\newcommand{\nat}{Nature}
\newcommand{\aap}{Astron. Astrophys.}
\def\apss{Astrophys. Space Sci.}
\def\physrep{Phys. Rep.}
\firstpage{1} 
\pubvolume{5}
\issuenum{5}
\articlenumber{125}
\pubyear{2018}
\copyrightyear{2018}
\history{Received: 21 March 2019; Accepted: 10 May 2019; Published: 22 May 2019}
 \updates{yes}
\Title{Fifty Years of Energy Extraction from Rotating Black Hole: Revisiting Magnetic Penrose Process $^\dagger$}

\Author{Arman Tursunov $^{1,2,3}$  and  Naresh Dadhich $^{4,*}$ }

\AuthorNames{Naresh Dadhich and Arman Tursunov}

\address{%
$^{1}$ \quad Institute of Physics and Research Centre of Theoretical Physics and Astrophysics, Faculty of Philosophy and Science, Silesian University in Opava, Bezru{\v c}ovo n{\'a}m.13, CZ-74601 Opava, Czech Republic; arman.tursunov@fpf.slu.cz\\
$^{2}$ \quad 1st Institute of Physics, University of Cologne, Z\"ulpicher Strasse 77, D-50937 Cologne, Germany \\
$^{3}$ \quad Ulugh Begh Astronomical Institute, Astronomicheskaya 33, Tashkent 100052, Uzbekistan \\
$^{4}$ \quad IUCAA, Post Bag 4, Ganeshkhind, Pune 411 007, India; nkd@iucaa.in \\
* \quad Correspondence: nkd@iucaa.in \\
$\dagger$ \quad This work is dedicated to the fond memory of Dr. Sanjay Wagh, who was one of the initiators of work on magnetic Penrose process in his doctoral thesis in mid 1980s. He passed away following a massive cardiac arrest in the morning of 15th May 2019,  the day this paper appeared on the arXiv. \\
}

\abstract{
Magnetic Penrose process (MPP) is not only the most exciting and fascinating  process mining the  rotational energy of black hole but it is also the favored astrophysically viable mechanism for high energy sources and phenomena. It operates in three regimes of efficiency, namely low, moderate and ultra, depending on the magnetization and charging of spinning black holes in astrophysical setting. In this paper, we revisit MPP with a comprehensive discussion of its physics in different regimes, and compare its operation with other competing mechanisms. We show that MPP could in principle foot the bill for powering engine of such phenomena as ultra-high-energy cosmic rays, relativistic jets, fast radio bursts, quasars, AGNs,    etc. Further, it also leads to a number of important observable predictions. 
All this beautifully bears out the promise of a new vista of energy powerhouse heralded by Roger Penrose half a century ago through this process, and it has today risen in its magnetically empowered version of mid 1980s 
from a purely thought experiment of academic interest to a realistic powering mechanism  for various high-energy astrophysical phenomena. }


\begin{document}
%%%%%%%%%%%%%%%%%%%%%%%%%%%%%%%%%%%%%%%%%%
%% Only for the journal Gels: Please place the Experimental Section after the Conclusions

\section{Introduction}

Among the experimental tests successfully passed by Einstein's general theory of relativity \cred{are the} tests of precession of Mercury's perihelion, deflection of photons by Sun's gravity, measurement of gravitational redshift, orbital decay of binary pulsars \cite{RevModPhys.66.699}, direct detection of gravitational waves \cite{PhysRevLett.116.061102}, investigation of properties of the Galactic center supermassive black hole \cite{Eckart-etal:2017:FOP:}, and  others. Thus  far, all experimental tests of general relativity at various scales and regimes bear no convincing evidence of any deviation of black holes from the rotating Kerr black hole hypothesis, which pronounces that an  astrophysical black hole can be well characterized solely by two parameters, namely its mass $M$ and spin $a$. 

The mass is the most fundamental parameter of a black hole, which in many cases can be measured with relatively high precision through observations of dynamics of nearby objects. For example, the~most recent estimates of the mass of order $4.14 \times 10^{6}~M_{\odot}$ of the currently best known black hole candidate SgrA* located at the center of our Galaxy have been achieved by near infrared  observations of S2 star revolving around central black hole \cite{2018A&A...615L..15G}.     To measure spin of a black hole one needs to probe effects occurring in strong gravity regime, as {\cred{its gravitational contribution has no Newtonian analog and hence it is very small and hard to measure}. Acceleration of interstellar matter floating towards a black hole \cred{gets heated  up}, resulting in X-ray emission from accretion disk or hot spots. Some progress on the spin determination methods has been achieved with observations and modeling of X-ray spectra from both stellar mass and supermassive black holes \cite{2011CQGra..28k4009M}.  In addition, potential detection of gravitational waves from extreme mass ratio inspirals (EMRIs) by future space-based Laser Interferometer Space Antenna (LISA) \cite{2017PhRvD..95j3012B} seems to be a  promising avenue for determination of spin of astrophysical black holes. However, since \cred{method of observation is model dependent and it cannot be directly measured, the estimated spin values} in various models may differ dramatically. 

According to the no-hair theorem, there can exist the third black hole parameter, electric charge arising from Einstein--Maxwell equations for rotating charged mass. The charge parameter of a black hole is usually set explicitly equal to zero, which is justified by quick discharge of any net charge of a black hole due to selective accretion of a plasma matter surrounding any astrophysical black hole. However, as black holes are usually embedded into external magnetic field arising due to \cred{plasma dynamics, and more specifically twisting of magnetic field lines due to the frame dragging of effect in the vicinity of a rotating black hole induces electric field in both vacuum and plasma cases  \cite{Wald:1974:PHYSR4:,Bla-Zna:1977:MNRAS:}. It posits a net quadrupole  charge on the  black hole \cite{1976Natur.264..525H,2018MNRAS.480.4408Z}.} This charge is weak in the same sense as magnetic field, i.e., its stress-energy tensor does not alter spacetime metric. Thus, the assumption of the Kerr hypothesis is \cred{well founded}. However, \cred it would turn out, as we    show below, that black hole charge would  play crucial role in making energy extraction process ultra-efficient, so much so that efficiency could range over \cblue{$10^{10}$}. In the next section, we   describe the black hole charging mechanisms in astrophysical context and discuss its possible screening by surrounding plasma. 

Spacetime around an astrophysical black hole is described by the Kerr metric in the standard form given in the Boyer--Lindquist coordinates, 
\bea
 d s^2 = - \left( 1- \frac{2Mr}{\RS} \right) d t^2 - \frac{4Mra \sin^2\theta}{\RS} \, dt d\phi  + \frac{\RS}{\Delta} \, d r^2 \nonumber\\
  + \RS\, d\theta^2 + \left( r^2 +a^2 + \frac{2Mra^2}{\RS} \sin^2\theta \right) \sin^2\theta \, d \phi^2,
 \label{KerrMetric}
\eea
where $\RS = r^2 + a^2 \cos^2\theta$ and $\Delta = r^2 - 2Mr + a^2$. Physical singularity occurs at the ring $r=0, \, \theta = \pi/2$. 

\cred{The roots of $\Delta = 0$ define outer and inner horizons located at $r_{\pm} = M \pm (M^2-a^2)^{1/2}$. It is the outer horizon which is referred as the event horizon $r_+ \equiv r_H$. The geometry is characterized by existence of the two Killing vectors, timelike, $\delta/\delta t$ and spacelike, $\delta/\delta \phi$, indicating the corresponding conserved quantities, energy $E$ and angular momentum $L$.} 
\cgreen{One can introduce an observer with timelike four-velocity and zero angular momentum $L=u_\phi = 0$, which is infalling into the black hole from rest at infinity.
This corresponds to the locally non-rotating frame of reference (LNRF) of the zero angular momentum observers (ZAMO) with the  four-velocity given by }
\bea \label{uZAMO}
 n^{\alpha} = (n^t,0,0,n^\phi),  \quad
 (n^t)^2=\frac{g_{\phi\phi}}{g_{t\phi}^2-g_{tt}g_{\phi\phi}}, \quad n^\phi=-\frac{g_{t\phi}}{g_{\phi\phi}}\,n^t.
\eea
 
 Computing the angular velocity of LNRF/ZAMO we get
 \beq 
 \Omega_{\rm LNRF} = - \frac{g_{t\phi} }{g_{\phi\phi}} = \frac{2 M a r}{(r^2 + a^2)^2 - a^2 \Delta \sin^2\theta}. 
 \eeq 
 
 Since $\Omega/(2 M a)$ is always positive, ZAMO co-rotates with the black hole being dragged.  }

One of the most interesting properties of the Kerr black hole geometry is the existence of direct analogy between event horizon area $A_H$ of a black hole with thermodynamical entropy $S_H$ (see, e.g.,~\cite{Bar-Car-Haw:1973:CMP:,Jacobson:1995:PRL:,Padmanabhan:2010:RRP:}), which  implies \cred {that a black hole of mass $M$ and spin $a$ has  irreducible energy,}
\beq
E_{\rm irr} = \sqrt{\frac{S_H \hbar c^5}{4\pi G k_{\rm B}}} \equiv \sqrt{\frac{A_H}{16\pi G^2}} c^4 = \frac{M c^2}{\sqrt{2}}\left[1+\sqrt{1-\left(\frac{a}{M}\right)^2}\right]^{\frac{1}{2}}.
\label{irrenegen}
\eeq

 For an extremely rotating black hole, this is 
 71\% of its total energy \cite{1971PhRvD...4.3552C,Bekenstein:1973:PHYSR4:}, while the rest of $29\%$ is the rotational energy and is thus available for extraction. For stellar mass black holes this energy is of order $10^{63}$ eV, while for supermassive black holes of mass $M=10^9~M_{\odot}$ it  is of order $10^{74}$~eV making them the largest energy reservoirs in the Universe. It is therefore most pertinent to tap this enormous source most effectively and ultra efficiently. \cred{In this paper, we   entirely address to classical black holes and their physics and astrophysics without any reference to }quantum effects, such as the Hawking radiation,    etc.

 The first attempt to tap energy from black hole was made by Roger Penrose in 1969  \cite{Penrose:1969:NCRS:} who pointed out existence of negative energy states of particles orbiting black hole with respect to observer at infinity. Negative energy orbits for neutral particles can exist inside the ergosphere. \cred{Static} observer with timelike trajectory, having spatial velocity { $\bf v = 0$}, 
 needs to satisfy the inequality $g_{tt} u^t u^t<0$ implying $g_{tt}<0$. \cred{This means static observers cannot exist when $g_{tt}$ turns positive; i.e., they can only exist} for  $r>r_{\rm stat}(\theta) \equiv M + (M^2-a^2 \cos^2\theta)^{1/2}$. The ergosphere is the region bounded by event horizon,  $r_H$ and static surface $r_{\rm stat}(\theta)$, so that $r_{\rm stat}(\theta) \geq r_H$. Energy of a particle with momentum $p_\alpha$ measured by an observer of velocity $u^\alpha_{(obs)}$ is $E=-p_\alpha u^\alpha_{(obs)}$. \cred{Since no observer can remain static below $r_{\rm stat}(\theta)$, i.e., $u^\alpha$ turns spacelike,  %implying 
 energy $E$ relative to an observer at infinity can turn negative for some suitable particle parameters. However,   local energy would now be defined relative to stationary observer---locally non-rotating or zero angular momentum observer---that has radial velocity zero but angular velocity is necessarily non-zero, $\omega=-g_{t\phi}/g_{\phi\phi}$, the frame dragging velocity. Thus, energy relative to LNRO/ZAMO is conserved and would however  be always positive while it could be negative relative to  observer at infinity. 
 Thus, in ergosphere  where $g_{tt}>0$, there exist particle orbits of negative energy states relative to infinity. This is the key property that drives Penrose process of energy extraction from a rotating black hole.} 

 Following the original idea of Penrose, let us consider the equatorial motion of a freely falling Particle 1 which decays inside the ergosphere into two  fragments one of which (Particle 2) attains negative energy relative to  infinity, while the other one (Particle 3) escapes to infinity with energy grater than that of incident particle. The efficiency of the process, defined as ratio of extracted to infalling  energy, and is given by the relation { (see, e.g.,    \cite{Wald:1974:APJ:,Bar-Pre-Teu:1972:APJ:} and derivation in Section \ref{mpp-regimes})}
 \beq \label{eff-pp}
 \eta_{\rm PP} = \frac{E_3 - E_1}{E_1} = \frac{1}{2} \left(\sqrt{\frac{2 M}{r}}-1\right).
 \eeq
 
 For split occurring close to horizon, it is then given by  
 \beq \label{eff-pp-rh}
 \eta_{\rm PP} = \frac{M}{2 a} 
 \left( \sqrt{2} \sqrt{ 1 - \sqrt{1 - \frac{a^2}{M^2}} } - \frac{a}{M} \right). 
 \eeq
 
 It is maximal for extremely rotating black hole ($a=M$), being  $\eta_{\rm PP} = 0.207$, or $\approx$21\%. For~moderate spins, e.g., $a=0.5 M$, PP efficiency is <$2\%$. In addition to low efficiency, PP is inoperable in realistic conditions  \cred{because threshold relative velocity between two fragments after split required to be greater than half of the speed of light  \cite{Bar-Pre-Teu:1972:APJ:,Wald:1974:APJ:}. This is the condition for a particle to ride on a negative energy orbit relative to an  observer at infinity. There exists no conceivable astrophysical process that could almost instantaneously  accelerate neutral particle to such  high velocities. Nor there exists any observational evidence to support such a happening. Thus, PP though very novel and purely relativistic in nature cannot be astrophysically viable.}

 In the mid-1980s, PP was revived astrophysically by inclusion of interaction of matter with electromagnetic field surrounding black hole  \cite{Wag-Dhu-Dad:1985:APJ:,Bha-Dhu-Dad:1985:JAPA:,Par-etal:1986:APJ:}, as nicely reviewed in \cite{Wagh-Dadhich:1989:PR:}. \cred{This was  magnetic version of  Penrose process (MPP), where the inconvenient relative velocity threshold between two fragments after split 
 could be easily overcome in presence of an external magnetic field in which black hole is immersed. In other words, energy required for a particle to get onto negative energy orbit could now come from particle's interaction with electromagnetic field leaving mechanical velocity completely free.}  It was then shown that the process turned very efficient and its efficiency could even exceed $100\%$. 
  For example, for electrons around stellar mass black hole, efficiency is greater than 100\%, for as low a field as  milliGauss \cite{2018MNRAS.478L..89D}. 
  Although energy extraction efficiency $100\%$ was first shown  in 1985  \cite{Wag-Dhu-Dad:1985:APJ:} for discrete particle accretion and idealized magnetosphere, it is highly gratifying to see that recent fully general relativistic magnetohydrodynamic (GRMHD) simulations  \cite{Nar-McC-Tch:book:2014:,Tch-Nar-McK:2011a:MNRAS:} have wonderfully borne out this most important and interesting feature of the process. 

{ Earlier in 1975, Remo Ruffini and James H. Wilson   considered the  process of energy extraction from rotating black hole based on the charge separation in a magnetized plasma accreting into Kerr black hole \cite{1975PhRvD..12.2959R}. This became the first use of gravitationally induced charge of a black hole by the frame-dragging effect for the extraction of rotational energy from black holes.}
  Another similar process of energy extraction from rotating black hole is the Blandford--Znajek mechanism (BZ) \cite{Bla-Zna:1977:MNRAS:}. BZ operates on the principle of unipolar generator, similar to classical Faraday disc. Here,   the role of disc is played by a black hole rotating in magnetic field.
  As in the case of MPP,  black hole's rotation generates electric currents along the horizon surface which convert mechanical spin energy of a black hole into electromagnetic energy to be  extracted. In addition, both BZ and MPP act due to existence of  quadrupole electric field, being produced by twisting of %
  magnetic field lines. However,  
  BZ requires force-free magnetosphere which can be formed e.g., by a cascade of electron--positron pairs \cite{Williams:2004:ApJ:}. This~requires the threshold magnetic field of order $10^4$  G. In addition, BZ cannot provide ultra high efficiency, which is the distinguishing feature of MPP  \cite{2018MNRAS.478L..89D}. {Below, we   show that MPP is a general process which includes and approximates to BZ for high magnetic field regime. As mentioned before, MPP operates in three regimes of efficiency:     low, moderate and ultra high. BZ is included in the middle---moderate efficiency regime. MPP has thus become one  of the leading processes for powering the % 
  central engine of high energy astrophysical objects such as quasars and AGNs involving black~holes.}
  
 Following the work of Roger Penrose, several other modifications and variants  of the original PP were proposed. These included  collisional Penrose process (CPP)  \cite{Pir-Sha-Kat:1975:APJL:,Pir-Sha:1977:APJ:} and its  various variants~\mbox{\cite{Zaslavskii:2016:PRD:,Ber-Bri-Car:2015:PRL:}} based on multiple collisions of particles within the ergosphere, whose energy in the center-of mass could  grow arbitrarily high for extremely rotating black hole (see for a recent review  \cite{Schnittman2018} and references therein). It was however agreed that CPP was unlikely to be relevant in realistic high-energy phenomena, since the efficiency of the process in astrophysical situations (i.e.,    non-extremal black hole, collision occurring not exactly on horizon, incident particle falls from infinity) was severely constrained with maximum $\eta_{\rm CPP} < 15$  \cite{2016PhRvD..93d3015L}. In \cite{1980BAICz..31..129S}, it has been shown that the efficiency of energy extraction from Kerr naked singularity can reach $157\%$. Electromagnetic fields and energy extraction from boosted black holes, i.e.,~black holes moving at relativistic speeds, have been studied in \cite{2014:MNRAS:MRA:,2015PhRvD..91h4044P}. Among similar high-energy  processes which have drawn attention in the literature is also the so-called BSW mechanism \cite{2009PhRvL.103k1102B}, which can provide arbitrarily high center-of-mass energy for collision again occurring at the horizon of extremely spinning black hole. It has also been generalized to include magnetic field  \cite{Fro:2012:PHYSR4:} and to many other cases of modified gravity (see, e.g.,      \cite{2015EPJC...75..399A,2015Ap&SS.360...19T,2016Ap&SS.361..288R,2014PhRvD..89j4048S,2013PhRvD..88l4001T,2013Ap&SS.343..173A,2013PhRvD..88h4036A,2013PhRvD..88b4016S,2011:PRD:AAA:}, among others).

\cgreen{ This paper is organized as follows. In {Section   \ref{sec2}}, we discuss the mechanisms of magnetization and charging of a rotating black hole in astrophysical context. In Section    \ref{sec3}, we describe the formalism of MPP, study its operation in different regimes and estimate the efficiency of MPP in several plausible radioactive decay modes. Here,   we show that the efficiency in some cases can exceed $10^{10}$ in astrophysically reasonable conditions. In Section   \ref{sec4}, we relate MPP for the explanation of ultra-high-energy cosmic rays (UHECRs) and relativistic jets. We show that the cosmic rays with the highest detected energy in the {range} $10^{18}$--$10^{20}$~eV can be explained by neutron beta-decay in ergosphere of SMBH of mass $10^9~M_\odot$ and magnetic field $10^4$  G. MPP can also provide possible explanation of the knee of the cosmic ray spectrum as the energy of proton after beta-decay in the ergosphere of Galactic center black hole reaches $\sim$$10^{15.5}$~eV. We also discuss how MPP can be related to the explanation of relativistic jets from black holes and AGNs. Employing different processes, such as the charge separation in a plasma, pair production, ionization of accretion disk and chaotic scattering, we discuss the jet-like motion of a matter, in particular the possibilities to obtain high Lorentz factors and strong collimation.  In Section \ref{sec-discussion},  we summarize the  main results and give concluding remarks.}

\section{Electromagnetized Black Hole} \label{sec2}
\unskip
\subsection{Magnetization of Black Holes}

Black holes are indeed embedded into external magnetic fields which can arise due to dynamics of surrounding plasma, e.g., electric currents inside accretion disks, intergalactic or  interstellar plasma dynamics. Magnetic field could also be generated in early phases of  expansion of the Universe. Strength of magnetic field can vary for each particular black hole candidate, usually being in the range $10$--$10^8$ G, highly dependent on properties of surrounding plasma. For example, the best known black hole candidate Sgr~A*, which is located at the Galactic center, is surrounded by highly oriented magnetic field measured at distance of few Schwarzschild radii from the center  has field of order $10$--$100$\,G  \cite{2015llg..book..391M,Eatough-etal:2013:Natur:}. Recent detection of orbiting hot-spots around the object \cite{2018A&A...618L..10G} indicates presence of strong poloidal magnetic field at the ISCO scale.

It is important to note that any astrophysical magnetic field is weak in a sense that its energy-momentum tensor does not modify the background Kerr metric. It is easy to see by  comparing magnetic field energy in a given volume with that of black hole's mass energy. This condition for stellar mass black holes   \cite{1978JETP...47..419G} is given by 
\beq
B \ll B_{\rm G} =  \frac{c^4}{G^{3/2} M_{\odot}}\left(\frac{M_{\odot}}{M}\right)\sim 10^{18} \frac{10 M_{\odot}}{M}\mbox{G}\, .
\label{BBB}
\eeq

Observations of various black hole candidates and astrophysical phenomena occurring in their vicinity indicate that the inequality in Equation (\ref{BBB}) is perfectly satisfied. This implies that an  astrophysical black hole is weakly magnetized, hence its effect on neutral test particle  dynamics is negligible. On the other hand, its effect for motion of charged particles is  non-ignorable---rather immense---as ratio of Lorentz to gravitational force would be very large due to large value of charge to mass ratio. Since matter surrounding black hole is usually moving with relativistic velocities and highly ionized, one can characterize relative influence of Lorentz to gravitational force by dimensionless parameter ${\cal B} = |q| B M/ (m c^4)$. For electrons close to event horizon of the Galactic center black hole  \cite{Eckart-etal:2017:FOP:}, the estimate of this parameter is 
\beq
\label{estimation-BBsgra}
{\cal B}_{\rm SgrA^*} \approx 2 \times 10^9 \left( \frac{q}{e} \right) \left( \frac{m}{m_{\rm{e}}} \right)^{-1} \left( \frac{ B } {10\,{\rm G }}\right) \left( \frac{M}{4 \times 10^{6}\,{M}_{\odot} } \right).
\eeq

For protons, this ratio is $\sim$$2000$ times lower. Stellar mass black holes, e.g., in binary systems can attribute magnetic fields of order $10^8$  G  \cite{2015ASPC..494..114P}, for which ${\cal B}$ is of similar order of magnitude as Equation~(\ref{estimation-BBsgra}). This implies that the effect of magnetic field on dynamics of charged particles is very dominant in realistic  astrophysical conditions.

In the Kerr geometry, it is natural to assume that external magnetic field would also share  symmetries of stationarity and axial symmetry. Using the Killing equation $\xi_{\alpha;\beta} + \xi_{\beta;\alpha} = 0$, one finds solution for electromagnetic field in the form \cite{Wald:1984:book:}
\beq \label{Amu-wald}
	A^{\mu} = C_1 \xi^{\mu}_{(t)} + C_2 \xi^{\mu}_{(\phi)}.
\eeq

The first solution of Maxwell equations in background rotating black hole spacetime corresponding to black hole embedded in homogeneous magnetic field aligned with the spin axis was obtained by R. Wald in \cite{Wald:1974:PHYSR4:}, while for  magnetic field inclined to the axis of rotation was due to J. Bi{\v c}{\'a}k and V. Jani{\v s} \cite{1985MNRAS.212..899B}. Magnetic field of current loop around rotating black hole corresponding to dipole magnetic field configurations
 was solved by J. Petterson in \cite{Petterson:1974:PHYSR4:}.

As we shown below, the orientation of magnetic field configuration plays rather secondary role in the context of MPP operation and its efficiency.
 The four-potential of asymptotically uniform magnetic field in which the Kerr black hole is immersed reads \cite{Wald:1974:PHYSR4:}
\bea
A_t &=& a B \left( \frac{M r}{\Sigma}
\left(1+\cos^2\theta\right) - 1 \right), \label{VecPotT} \\
%\eeq
%\beq 
A_{\phi} &=& \frac{B}{2} \left(r^2+a^2-\frac{2 M r a^2}{\Sigma}
\left(1+\cos^2\theta\right)\right) \sin^2\theta.
\label{VecPotP}
\eea

One can notice that rotation of black hole generates quadrupole electric field given by $A_t$ which is the result of twisting of magnetic field lines---the frame-dragging effect.  Electric field due to this induced charge is for arbitrary magnetic field configuration in the vicinity of axially symmetric black hole. This 
induced quadrupolar charge may be referred as  black hole charge    \cite{2018MNRAS.480.4408Z,2018arXiv180807887L} and it is this which is responsible for  providing necessary negative energy to particle in the ergosphere. By this way, the inconvenient and unsurmountable condition on relative velocity for the original mechanical PP could be easily overcome. 
One should note that above is true until the extremity of black hole spin is reached, where the gravitational analog of Meissner effect comes into play. Extremely rotating black hole expels fields out,  thus acting as a superconductor  \cite{Wald:1974:PHYSR4:,Bic-Led:2000:NCBS:},  which happens exactly at $a/M=1$. On~the other hand, it is generally assumed that the maximum plausible spin of astrophysical black holes is $a/M = 0.998$ \cite{Thorne:1974:ApJ:} marking equipartition of magnetic and gravitational energy. Thus, the~further study and discussion is well in order and well motivated.

\subsection{Black Hole Charge}

Assumption of electrical neutrality of a black hole in many cases is justified by the presence of a plasma around black hole which can quickly discharge any charge excess. 
Indeed, one can estimate the discharge timescale of maximally charged black hole using the following arguments.
Maximal theoretical value of the charge of a black hole of mass $M$ is given by $Q = 2 G^{1/2} M$, written in Gaussian units. For spinning black hole, it     is of order  \cite{2018MNRAS.480.4408Z}
\begin{equation}
    Q_{\rm max} \approx 3.42 \times 10^{20} \left(\frac{M}{M_{\odot}}\right) {\rm C}.
\end{equation}

If such a charge is carried by protons and electrons, this corresponds to the  mass $M_Q = m_{ p,e} Q /e$, where $p$ and $e$ denote proton and electron, respectively. 
Luminosity of black hole surrounded by  plasma or accretion disk can be derived from infalling matter as $L = \epsilon \dot{M} c^2$, where $\dot{M}$ accretion rate and $\epsilon$ is the fraction of the rest mass energy radiated away. On the other hand, from the balance of gravitational force and radiation pressure in the vicinity of a black hole one can derive the Eddington luminosity for fully ionized hydrogen plasma surrounding a black hole in the form
\begin{equation}
    L_{\rm Edd} = \frac{4\pi G M m_p c}{\sigma_{\rm T}} \approx 1.26 \times 10^{38} \left(\frac{M}{M_{\odot}}\right) {\rm erg/s}.
\end{equation}

Defining  charged matter accretion rate as the fraction of total accretion rate, $\dot{M}_Q = \delta \cdot \dot{M}$, we get the neutralization timescale of the maximally charged black hole as 
\begin{equation}
t_{Q, {\rm acc.}} = \frac{4}{3} \frac{e^3 \, \epsilon \, m_{p,e} }{ G^{1/2} \,c^3 \, \delta \,  m_p m_e^2} \approx 2.5 \times 10^{-2} \left( \frac{m_{p,e}}{m_p} \right) \left(\frac{\epsilon}{\delta}\right) {\rm s}.
\end{equation}
which is estimated for positive black hole charge. In the case of negative charge, the timescale is $\approx$$1835$ times lower. Both ${\epsilon}$ and $\delta$ have values in the range $(0\, , \, 1)$ and in many cases are of similar order of magnitude. This implies that in all  astrophysically relevant settings any net charge on black hole would be neutralized relatively  quickly unless there is a mechanism preventing the black hole from neutralization.

{
For a spinning black hole immersed in an external stationary magnetic field, it is
easy to see from Equations (\ref{Amu-wald}) and (\ref{VecPotT})   that any local observer within the ergosphere can measure  nonzero
electric field component if the magnetic field has nonvanishing poloidal component  \cite{Wald:1974:PHYSR4:,1975PhRvD..12.2959R,2009JKPS...54.2503K}. 
 Black hole spin contributes to the Faraday induction generating the electrostatic potential $A_t$,
which can be associated with induced electric field. The charging process is similar to the classical
Faraday's homopolar generator. In case of the uniform external magnetic field, there exists the
potential difference %
between event horizon and infinity taking the form 
\beq
\Delta \phi = \phi_{\rm H} - \phi_{\infty} = \frac{Q - 2 a M B}{2 M},
\eeq
which leads to the % 
selective accretion %
into black hole % 
until the potential %
difference vanishes. In that
case the black hole acquires net charge $Q= 2 a M B$. The process of charging of a black hole due to
spin-induced effect in magnetic field takes place in any magnetic field configuration which shares the
symmetry of background Kerr metric spacetime. On the other hand, the chosen form of the field configuration
may put restrictions on the dynamical timescales of the process of selective accretion due to following
charge separation in a plasma. In general, the energy of charged test particle is given by $E=-P_{\mu}
\xi^{\mu}_{(t)} = - (m u_{\mu} - q A_{\mu}) \xi^{\mu}_{(t)}$. Difference between electrostatic energy %
 of a charged particle at the event horizon and at infinity is given by $E_H - E_{\infty} = q
 A_{t|r\rightarrow r_H} - q A_{t|r\rightarrow \infty} \equiv \delta$. For positive $\delta$, more
 favorable is the accretion of particles with the sign of charge of $q$, while for negative $\delta$,
 it is more likely to accrete particles with $-q$ sign of charge.
 In both cases, this leads to the formation of the net black hole charge with a sign which depends on
 the orientation of magnetic field with respect to the rotational axis of a black hole
 \cite{Tur-Stu-Kol:2016:PHYSR4:}. For a magnetic field generated by the dynamics of co-rotating
 surrounding plasma matter, the black hole's charge is more likely to be positive.
In \cite{2018MNRAS.480.4408Z}, it  is shown that, in realistic cases applied to the Galactic center black
hole, even a small charge of the black hole can have non-negligible effects on the observed
bremsstrahlung emission profile.
We would like to emphasize that the black hole charge plays a key role in the black hole energy
extraction processes, such as Blandford--Znajek mechanism  and magnetic Penrose process
\cite{2018MNRAS.478L..89D}. Discharge of induced electric field by oppositely charged accreting matter
drives the rotational energy away from the black hole in both of the processes.    
}

However,   to tap the gravitationally induced electrostatic energy from black hole and support particle acceleration, this electric field should not be screened as usually occurs in the presence of a plasma. In the past the problem has been widely discussed  \cite{2001bhgh.book.....P,1997PhyU...40..659B} resulting in the Komissarov's theorem \cite{2004MNRAS.350..427K}, which states that induced electric field of rotating black hole in magnetic field is not screened at least within ergosphere. Indeed, total screening of black hole's charge may occur only when the following two conditions are satisfied simultaneously
\beq \label{Kom-theor}
\vec{B} \cdot \vec{D} = 0, \quad B^2 - D^2 >0,
\eeq 
where $B$ and $D$ are magnetic and electric field measured LNRO or ZAMO. It is easy to show that the relation $B^2 - D^2$ which is positive far away from black hole, is negative everywhere inside the ergosphere \cred{\footnote{As for energy $E$, sign of $B^2-D^2$ would be dependent upon the location of observer, whether inside or outside ergosphere. There could exist no static observer in ergosphere, it could at best be stationary with non-zero frame dragging angular velocity.}}    \cite{2004MNRAS.350..427K}. Moreover,  the sign of this relation is independent of the strength of magnetic field being only dependent on the location. In other words, electric field is stronger for stronger magnetic field and within the ergosphere it cannot be screened off. 

Black hole charge associated with gravitationally induced electric field has been estimated in \cite{2018MNRAS.480.4408Z} for the Galactic center supermassive black hole Sgr~A*, with an upper limit of $10^{15}$ C. On the other hand, classical estimates of charge, based on the difference between thermal velocities and masses of electrons and protons %
in the fully ionized plasma around Sgr~A*, imply the presence of equilibrium charge of the central body of order $10^8$ C. It therefore follows that  black holes posses an electric charge in the range $10^2$--$10^{12}$ C per solar mass. This charge is gravitationally weak in a sense that its influence to the spacetime metric can be neglected. \cred{For it to be gravitationally significant would require charge of order $\sim$$10^{20}$ C per solar mass, and, therefore, Kerr black hole hypothesis stands firm and valid.} It~is important to note that such a charge cannot be measured by imaging of black holes, based on the observations of their shadows.
However, even such a small charge associated with the black hole has  %  
 significant effects on the processes occurring in its neighbourhood, such as %
 acceleration of charged particles to ultra-high energy.

\section{Magnetic Penrose Process} \label{sec3}
\unskip
\subsection{Dynamics of Charged Particles around Black Hole}

Efficiency of MPP in its original formulation can be derived by applying the conservation law for parameters of the particles before and after decay near black hole. Motion of relativistic charged particles, such as electrons and protons in magnetic field leads to inevitable synchrotron radiation loss which can sufficiently influence  particle's energy and angular momentum. Therefore,  it is important to establish the limits of applicability of the formalism based on the conservation laws. In this subsection, we provide general description of the charged particle  dynamics in curved spacetime taking into account the influence of radiation-reaction forces and providing corresponding estimates.

Detailed studies of charged particle dynamics around Schwarzschild and Kerr black holes in magnetic field neglecting radiation reaction were done by several authors (see, e.g.,    \cite{Fro-Sho:2010:PHYSR4:,Kol-Stu-Tur:2015:CQG:,Tur-Stu-Kol:2016:PHYSR4:,Kov-etal:2010:CLAQG:}). Qualitative and quantitative studies of synchrotron radiation reaction problem in curved background can be found, e.g., in  \cite{Sokolov-etal:1983:SovPJ:,Sok-Gal-Pet:1978:PLA:}, and in more recent papers  \cite{Shoom:2015:PHYSR4:,2018ApJ...861....2T}.
Equation of motion for a point charge in its most general form is usually referred as the DeWitt--Brehme equation \cite{DeW-Bre:1960:AnnPhys:}, which contains non-local tail integral term and Ricci term. Detailed discussion of the complete set of equations of motion for radiating charged particles can be found in  \cite{Hobbs:1968:AnnPhys:,Poisson:2004:LRR:}. It was shown, however, that the tail force can be neglected in most of the cases \cite{2018ApJ...861....2T}, being for electrons around stellar mass black hole $10^{-19}$ times smaller than corresponding gravitational ``force'' on the  horizon scale. Ricci term is irrelevant in vacuum metrics. Thus, after several algebraic manipulations and applying Landau--Lifshitz method~\cite{Lan-Lif:1975:CTF:}, one can simplify the equation of motion for radiating charged test particle around black hole to the following form 
 \cite{2018ApJ...861....2T}
\beq 
 \frac{D u^\mu}{\d \tau} = \frac{q}{m} F^{\mu}_{\,\,\,\nu} u^{\nu} 
+
\frac{2 q^3}{3 m^2}  \left(\frac{D F^{\alpha}_{\,\,\,\beta}}{d x^{\mu}} u^\beta u^\mu + \frac{q}{m} \left( F^{\alpha}_{\,\,\,\beta} 
F^{\beta}_{\,\,\,\mu} +  F_{\mu\nu} F^{\nu}_{\,\,\,\sigma} u^\sigma u^\alpha \right) u^\mu \right),
\label{eqmoLL}  
\eeq 
where $u_\alpha$ is four-velocity satisfying normalization $u^\alpha u_\alpha = -1$ and $F^{\mu}_{\,\,\,\nu}$ is the Faraday tensor of external electromagnetic field, whose covariant derivative is given by
\begin{equation}
\frac{D F^{\alpha}_{\,\,\,\beta}}{d x^{\mu}} = \frac{\partial F^{\alpha}_{\,\,\,\beta}}{\partial x^{\mu}} + \Gamma^{\alpha}_{\mu\nu} F^{\nu}_{\,\,\,\beta} - 
\Gamma^{\nu}_{\beta\mu} F^{\alpha}_{\,\,\,\nu}.
\end{equation}

Equation (\ref{eqmoLL}) is thus the covariant form of Landau--Lifshitz equation. Solving this equation numerically gives us the rate of energy and angular momentum loss. For our purpose here, it is important to find the cooling timescales corresponding to  energy and angular momentum loss of charged particles around black hole. 
Therefore, applied to the Schwarzschild black hole with uniform magnetic field, we find the evolution of specific energy ${\cal E}$ and specific angular momentum ${\cal L}$ in the~form
\bea \label{lossSCH}
\frac{d {\cal E}}{d\tau} &=& - 4 k \BB^2 \ce^3 + 2 k \BB \ce \left(2\BB f + \frac{u^{\phi}}{r} \right), \\
\frac{d {\cal L}}{d\tau} &=& 4 \BB^2 k u^{\phi} \left( f^2 (u^t)^2  - f \right) - 2 u^r u^{\phi} \left(r - 4 \BB^2 k^2 \right) + 2 r \BB u^r,\\
\BB &=& \frac{ q G B M}{m c^4}, \label{cBB}
\eea
where $f = 1-2M/r$ and $k = 2q^2/(3 m)$, \cred{and $q, m$ are mass and charge of test particle}.
 For~ultrarelativistic charged particle ($\ce \gg 1$, or $\BB \gg 1$), 
the leading contribution for energy loss is the first term on the right hand side of Equation    (\ref{lossSCH}). Inserting all constants, the cooling timescale of charged particle is given by
\beq \label{life-time}
\tau_{\rm cooling} \approx \left(1-\frac{2\, G M}{r \, c^2}\right)^{-1} \frac{3\, m^3 c^5}{2\, q^4 B^2}.
\eeq

In this case the Lorentz force is dominant over gravitational ``force'', which happens in realistic settings as discussed above  (Equation    (\ref{estimation-BBsgra})). Closer to black hole the cooling timescale increases. In~Table~\ref{tab1}, we give typical cooling timescales of electrons, proton and fully ionized iron nuclei for various values of magnetic field. Due to cubic dependence on mass, { electrons cool $10^{10}$ times faster than protons. }
One can compare the cooling timescale with an orbital timescales of particles at ISCO $\tau_{\rm orb} \approx 4 \pi r_{\rm isco}/c$, which is of order   $\sim$$10^{-3}$~s for  non-rotating stellar mass black holes or $\sim$$10^{5}$~s for supermassive black holes. Thus, the energy loss can be quite relevant, especially in the case of lighter particles, such as electrons.

%%%%%%%%%%%%%%%%%%%%%%%%%%%%%%%%%%%%%%%%%%%%%%%%%%%%%%%%%%%%%%%%%%%%%%%%%%%
\begin{table}[H]
\caption{Typical cooling times of electrons $\tau_e$, protons $\tau_p$ and fully ionized iron nucleus $\tau_{\rm Fe}$ for various values of magnetic field strength $B$.}
\label{tab1}
%\begin{center}
\centering
\begin{tabular}{c c c c  }
%\hline
\toprule
\textbf{B (Gauss)}    &  \boldmath$\tau_e$ \textbf{(s)}  &  \boldmath$\tau_p$ \textbf{(s) }& \boldmath$\tau_{\rm Fe}$ \textbf{(s)  } 	  \\	 										
%\hline \hline 
\midrule

$10^{12}$	   &  ~~~$10^{-16}$~~~ & ~~$10^{-6}$~~ & ~~$10^{-5}$~~   \\	
$10^{8}$	   &  ~~~$10^{-8}$~~~ & ~~$10^{2}$~~ & ~~$10^{3}$~~   \\				
$10^{4}$	   &  ~~~$1$~~~       & ~~~$10^{10}$~~~ & ~~$10^{11}$~~   \\				
$1$	         &  ~~~$10^{8}$~~~ & ~~~$10^{18}$~~~ & ~~$10^{19}$~~   \\
$10^{-4}$   &  ~~$10^{16}$~~   & ~~$10^{26}$~~  & ~~$10^{27}$~~    \\						

%\hline
\bottomrule
\end{tabular}

%\end{center}
\end{table}
%%%%%%%%%%%%%%%%%%%%%%%%%%%%%%%%%%%%%%%%%%%%%%%%%%%%%%%%%%%%%%%%%%%%%%%%%%%

If the radiation-reaction force is neglected, two components of generalized four-momentum $P_\alpha = m u_\alpha + q A_\alpha$ are conserved, namely energy and angular momentum of a particle. They can be associated with the Killing vectors in the form
\bea
- {\cal E} = \xi^{\mu}_{(t)} \frac{P_{\mu}}{m} = g_{tt} \frac{dt}{d\tau} + g_{t\phi}\frac{d\phi}{d\tau} + \frac{q}{m} A_{t},&& \label{Energy} \\ %
{\cal L} = \xi^{\mu}_{(\phi)} \frac{P_{\mu}}{m} = g_{\phi\phi} \frac{d\phi}{d\tau} + g_{t\phi}\frac{dt}{d\tau} + \frac{q}{m} A_{\phi}.&& \label{AngMom}
\eea

In this case, the standard approach can be used (see, e.g.,     \cite{Tur-Stu-Kol:2016:PHYSR4:} and references therein).

The motion of charged particle in magnetic field is always bounded, which can be described by introducing the effective potential. In the equatorial motion, it takes the form
\beq 
V_{\rm eff} = - \frac{q}{m} A_t  %
- \frac{g_{t\phi}}{g_{\phi\phi}} ({\cal L} - \frac{q}{m} A_{\phi}) 
+ \left[ 
\left(-g_{tt} - \omega g_{t\phi}\right) \left(\frac{({\cal L} - \frac{q}{m} A_{\phi})^2}{g_{\phi\phi}}+1\right) \right]^{1/2},
\label{Veff-mpp}
\eeq

One sees that $V_{\rm eff}$ can attain both positive and negative values depending on the angular momentum, charge and location of the particle.  Occurrence of negative energy orbits \cgreen{ (NEOs) in the ergosphere is critical requirement for extraction of energy from rotating black holes.  Presence of the term $-q A_t$ in the effective potential extends  the region of existence of NEOs far beyond the ergosphere, while the last two terms in $V_{\rm eff}$ can be negative only within the ergosphere. In fact,  the~region with possible NEOs and, thus, the energy extraction zone for charged test particles extends to infinity  \cite{1984PhRvD..29.2712D,1984PhRvD..30.1625D}.} 
 Ergosphere is maximal in the equatorial plane, therefore we consider decay of a particle falling onto black hole in the equatorial plane.

In addition to energy and angular momentum and their conservation due to the Killing symmetries,  the normalization condition $u^\alpha u_\alpha = - k$ must be satisfied for both charged and uncharged particles, where $k=1$ for massive particle and $k=0$ for massless particle. 
In the equatorial motion with the four-velocity $u^\alpha = u^t ( 1, v, 0, \Omega)$, where $v=dr/dt$ and $\Omega = d\phi/dt$, we get the angular velocity of a test particle with respect to the asymptotic observer at rest in the form
\bea \label{eq-Omega-gen}
\Omega &=& \frac{1}{B} \left( -C g_{t\phi}
\pm \sqrt{u_t^2 \left(C g^2-A g_{rr} v^2\right)} \right),\\
B &=& k g_{t\phi} + u_t^2 g_{\phi\phi}, \quad C = k g_{tt} + u_t^2, \\
g^2 &=&  g_{t\phi}^2-g_{\phi\phi} g_{tt}, \quad u_t = - \left( {\cal E} + q/m A_t \right).
\eea
where the sign defines the co or counter rotation with respect to LNRO.  
The limit of $u^\alpha$ tending to a null vector gives the restrictions to the angular velocity of a particle (both charged and uncharged) surrounding black hole in the form \cite{Par-etal:1986:APJ:}
\beq \label{eq-Omega-pm}
\Omega_{-} \leq \Omega \leq \Omega_{+}, \quad \Omega_{\pm} = \frac{1}{g_{\phi\phi}} \left(-g_{t\phi} \pm \sqrt{g_{t\phi}^2 - g_{tt} g_{\phi\phi}} \right).
\eeq 

\subsection{Split of Infalling Particle}

Let us now consider decay of a Particle 1, which is not necessarily neutral, into two charged %
 Fragments 2 and 3 close to horizon in the equatorial plane. According to conservation of energy and angular momentum after decay, we write
\begin{eqnarray}
E_1 &=& E_2 + E_3, \\
L_1 &=& L_2 + L_3, \\
q_1 &=& q_2 + q_3, \\
m_1 \dot{r}_1 &=& m_2 \dot{r}_2 + m_3 \dot{r}_3, \label{conserv-req} \\
0 &=& m_2 \dot{\theta}_2 + m_3 \dot{\theta}_3, \\
m_1 &\geq& m_2 + m_3,
\end{eqnarray}
where a dot indicates derivative relative to  particle's proper time.
If energy of Particle 2 is negative relative to infinity, then Particle 3 attains energy $E_3 = E_1 - E_2 > E_1$ greater than that of incident Particle 1. Infalling negative energy into the black hole results in extraction of its rotational energy. 
Using the conservation laws, one can show that the local angular velocities of particles at the point of split satisfy the relation
\beq \label{eqmuphi}
m_1 u^{\phi}_1 = m_2 u^{\phi}_2 + m_3 u^{\phi}_3.
\eeq 

Reminding that $u^\phi = \Omega ~u^t = - \Omega ~e/X$, where $e = (E+qA_t)/m$ and $X = g_{tt} + \Omega ~g_{t\phi}$, Equation~(\ref{eqmuphi}) can be rewritten  in the form
\beq 
\Omega_1 m_1 e_1 \frac{X_2 X_3}{X_1} = \Omega_2 m_2 e_2 X_3 + \Omega_3 m_3 e_3 X_2.
\eeq

After several algebraic manipulations, we get the energy of escaping particle in the form
\begin{eqnarray}
E_3 &=& \chi (E_1 + q_1 A_t) - q_3 A_t, \label{E3-mpp} \\
\chi &=& \frac{\Omega_1 - \Omega_2}{\Omega_3 - \Omega_2} \frac{X_3}{X_1}, \quad X_i = g_{tt} + \Omega_i g_{t\phi}. \label{chi-mpp}
\end{eqnarray}
where $\Omega_i = d\phi / d\tau$ is the angular velocity of $i$th particle, which is given by Equation    (\ref{eq-Omega-gen}) and restricted by the limiting values (Equation \eqref{eq-Omega-pm}). If~the parameters are chosen in such a way that at the point of split $q_3 A_t<0$, then this term plays the dominant role in the energy extraction from black hole. If~$q_3>0$, and $B$ and $a$ are positive, it is easy to see that the condition   $q_3 A_t<0$ is perfectly satisfied.

\subsection{Three Regimes of MPP} \label{mpp-regimes}

MPP can operate in three regimes providing low, moderate and ultra high  efficiency for the energy extraction from black hole, depending essentially on strength of magnetic field.
As in the neutral case, Equation    (\ref{eff-pp}), we define the energy extraction efficiency as the ratio between %
gain and input energies, i.e., in our notation
\beq  \label{eff-def}
\eta = \frac{E_{3} - E_{1}}{E_{1}} = \frac{- E_{2}}{E_{1}}.
\eeq

Using Equations  (\ref{E3-mpp}) and (\ref{chi-mpp}) at the point of split, the general expression for the efficiency reads
\begin{equation} 
\eta_{\rm MPP} = \chi - 1 + \frac{\chi ~ q_1 A_t - q_3 A_t}{E_1}, \label{efficMPP}
\end{equation}
where $A_t$ is calculated at the point of split. Setting $\Omega_1$ to $\Omega$ given by Equation    (\ref{eq-Omega-gen}) and  $\Omega_2 = \Omega_{-}$,  $\Omega_3 = \Omega_{+}$, which maximizes the efficiency, and reminding that the velocity component $u_t$ is related to energy $E$ as $m u_t = -(E+qA_t)$, we get 
\beq
\chi = \frac{1}{2} \left(1+ \sqrt{1+ \frac{k}{u_t^2}~g_{tt} }\right),
\eeq 
which for freely falling massive particle ($k=1$) reduces to $\chi = (1+\sqrt{1+g_{tt}})/2$. 
Equation  (\ref{efficMPP}) can also be rewritten as 
\begin{equation} 
\eta_{\rm MPP} = \eta_{\rm PP} + \frac{q_3 A_t - q_1 A_t (\eta_{\rm PP}+1)}{m_1 u_{t1} + q_1 A_t}, \label{efficMPP-PP}
\end{equation}
{ where 
\beq
\eta_{\rm PP} = \chi - 1 = \frac{1}{2} \left(\sqrt{1+g_{tt}})-1 \right),
\eeq 
is the efficiency of original Penrose process leading to Equations  (\ref{eff-pp}) and (\ref{eff-pp-rh}). } 
In the absence of magnetic field, MPP turns to original Penrose process giving its lower limit with the maximum efficiency $\eta^{\rm low}_{\rm MPP} \equiv \eta_{\rm PP} = (\sqrt{2} - 1)/2 \approx 0.207$ or $20.7 \%$, corresponding to  extremely rotating black hole. 
\pagebreak

If all particles are charged, the efficiency is given by its full form (Equation \eqref{efficMPP-PP}). However, as     shown above, in the presence of magnetic field for elementary particles the electromagnetic forces are dominating in the system, implying that $|\frac{q}{m} A_t| \gg |u_t|$. Relating this to Equation (\ref{efficMPP-PP}) one can simplify the expression to the form
\beq \label{MPP-mod}
\eta^{\rm mod.}_{\rm MPP} \approx \frac{q_3}{q_1} - 1.
\eeq 

This is the moderate regime of MPP, which can operate, as we see, when $q_3 > q_1$, thus neutralizing gravitationally induced electric field of black hole. MPP in moderate regime has direct analogy with another famous process, namely Blandford--Znajek mechanism (BZ). In both cases, the driving engine is a quadrupole electric field of a black hole which arises due to twisting of magnetic field lines by the frame-dragging effect.
It is important to note, that the efficiency of BZ and moderate regime of MPP cannot grow ultra-large due to natural restrictions of global plasma neutrality surrounding black hole. In moderate regime MPP approximates to BZ and that explains why ultra-high efficiency has not been observed in numerical simulations of the process.

There could as well exist the third and the most efficient regime of MPP, which requires special attention leading to several important predictions. 
{ If the Particle 1 is neutral $q_1 = 0$, with energy $E_1 = m_1$,  splitting into two charged fragments, it is easy to see that the general expression for efficiency  (Equation \eqref{efficMPP}) reduces to the form
\beq \label{MPP3}
\eta^{\rm ultra}_{\rm MPP} = \chi - 1 - \frac{q_3 A_t}{m_1} .
%= \frac{1}{2} \left(\sqrt{\frac{2 M}{r}}-1\right) - \frac{q_3 A_t}{E_1} 
%\approx \frac{q_3}{m_1} A_t. 
\eeq 

%The term $\chi - 1$ on the right hand side of (\ref{MPP3}) depends on purely geometrical factors with the possible values of $\chi - 1$ varying in the range ($0; 0.207$).
The term $\chi - 1$ on the right hand side of  Equation (\ref{MPP3}) is the efficiency of mechanical PP, which
depends on purely geometrical factors and its value ranges from $0$ for $a=0$ to $0.207$ for $a=1$. 
One can also see  that $A_t$ component of vector potential  (Equation \eqref{VecPotT}) attains negative values everywhere above horizon, if the spin and magnetic field are co-aligned, i.e., $a B > 0$.
Since for elementary particles, such as electrons and protons, the charge-to-mass ratios $q/m$ are typically very large, the dominant contribution in Equation  (\ref{MPP3}) is due to $-q_3 A_t/m_1$ term. Thus,
the expression for the efficiency in this case can be rewritten in the form
\beq \label{MPP4}
\eta^{\rm ultra}_{\rm MPP} \approx - \frac{q_3}{m_1} A_t,
\eeq 
which can grow to enormous values for elementary particles.  If the spin and magnetic field are aligned in opposite directions, i.e., $a B < 0$, then,   to get positive $\eta^{\rm ultra}_{\rm MPP}$, the charge of escaping particle must be negative, $q_3<0$.  } This causes MPP to turn ultra-high-efficient regime imparting ultra-high energy to escaping particle. 

Note  that all expressions for efficiency in three regimes are quite general and independent of magnetic field configuration. We also do not specify splitting point at this stage. \cred{Essentially axial symmetry of spacetime and electromagnetic field is what is required for MPP to operate in three regimes of efficiency depending upon the two parameters, magnetic field strength and charge to mass ratio of particles involved in the process of energy extraction. The other factor that matters is split point should be as close to horizon as possible which means infalling particle is neutral so that it can 
reach closer to horizon without any hindrance and then it splits or decays into charged fragments of opposite charge. Thus,   the ultra-high regime is characterized by magnetic field, particles involved are electron or protons, and it is neutral particle that decays closest to horizon. } 
\pagebreak

The field configuration however matters for direction of escape and  final fate of escaping particle, which is expected to move along the magnetic field lines. For example, in case of uniform magnetic field the collimation of escaping particles is maximal, while in case of hypothetical magnetic monopole the particles escape isotropically. In the case of closed magnetic field lines such as for dipole magnetic field generated by current loops in the accretion disk of black hole, the escaping zone is concentrated on the polar caps collimating bipolar outflow of matter similar to relativistic jets. On the other hand, if~energy of escaping particle is ultra-high powered by ultra-high efficient MPP,  particle can get across the field lines and escape  in arbitrary direction.

\subsection{Quantitative Estimates}

    To compare different regimes of MPP, we choose the external magnetic field to be asymptotically homogeneous \cite{Wald:1974:PHYSR4:} with the strength $B$. Thus, the four-vector potential $A^\mu$ around black hole is given by the relations in Equations (\ref{VecPotT}) and (\ref{VecPotP}). We start from the  expression for efficiency of MPP in ultra-high regime, given by
\beq \label{MPPeffic}
\eta_{\rm MPP}^{\rm ultra} = \frac{1}{2} \left(\sqrt{\frac{r_S}{r_{\rm split}}} - 1\right) + \frac{q_3 G B M \,a}{m_1 c^4} \left(1 - \frac{r_S}{2 r_{\rm split}}\right),
\eeq
where $B$ is the strength of uniform magnetic field, $r_S = 2 G M/c^2$, and $r_{\rm split}$ is the splitting point. The~first term on the right hand side is positive only if the splitting point $r_H \leq r_{\rm split} < r_S$ with maximum at $r_{\rm split} = r_H$. 
This, purely geometrical, term is due to the original Penrose process and it varies in the range $(0\, ; \, 0.21)$. The second term gives the contribution due to electromagnetic interaction, which can exceed $1$ for electrons with a few milliGauss field and stellar mass black hole \cite{2018MNRAS.478L..89D}.  \cred{The~important point to be noted is that electromagnetic interaction effectively expands ergosphere beyond the geometric bound, $r=2M$. That is, negative energy orbits are available in much enlarged region which goes on 
to enhance efficiency and overall working of the process.}

\begin{figure}[H]
  \centering
 \includegraphics[width=0.3\textwidth]{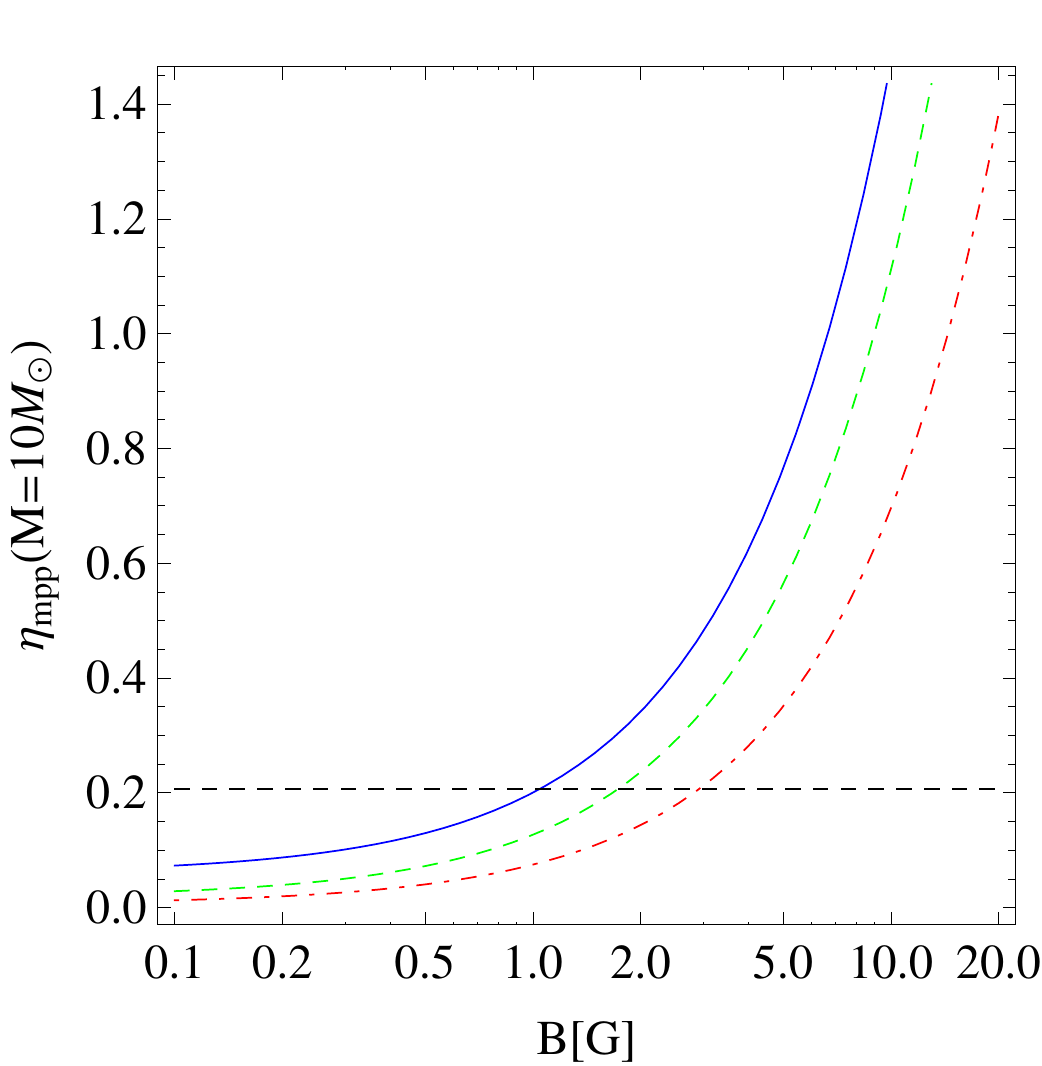}
 \includegraphics[width=0.3\textwidth]{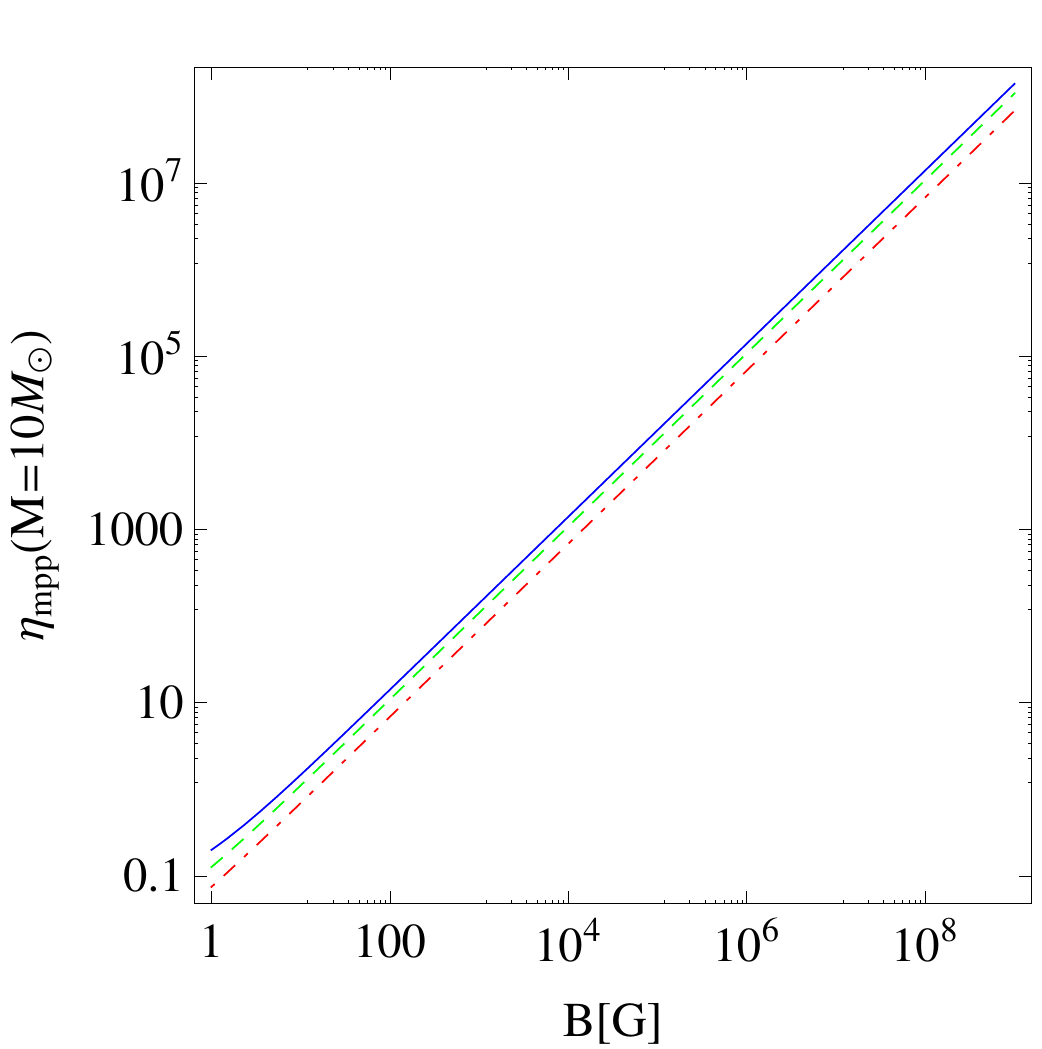}
 \includegraphics[width=0.3\textwidth]{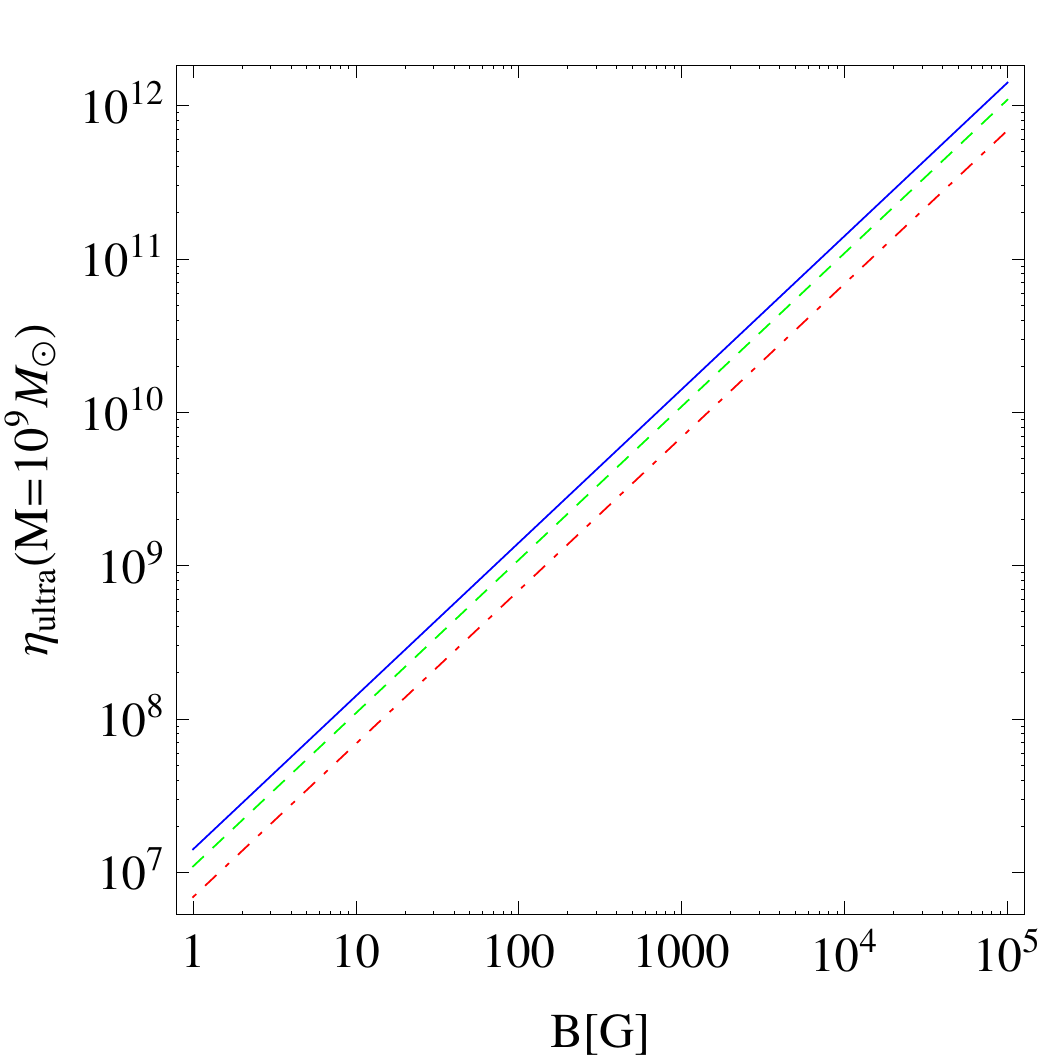}
  \caption{Efficiency of MPP in ultra-high regime: (\textbf{left},\textbf{middle})  stellar mass black hole of mass $10~M_\odot$; and (\textbf{right}) supermassive black hole of mass $10^9~M_\odot$. Dashed horizontal line on the left plot correspond to the efficiency of original Penrose process for black hole with extremal spin. Colored curves correspond to various spins: $a=0.8$--- 
  solid blue, $a=0.5$---dashed green and $a=0.3$---dot-dashed red.  }
  \label{fig-eff-ultra}
\end{figure}

 Efficiency of MPP for escaping proton in ultra-high regime after decay of freely falling neutron above horizon is plotted against magnetic field in Figure   \ref{fig-eff-ultra} for different values of black hole mass and spin. The interesting and important that emerges from this is that MPP does not require rapid rotation of black hole and can as well operate for low spins.
In Table~\ref{tab2}, we compare  MPP efficiency in different  regimes for some typical radioactive decay modes occurring above horizon of a black hole with mass 10 $M_\odot$, spin $a=0.8M$ and magnetic field of strength $10^4$G. 
 Magnetic field is taken to be asymptotically~uniform.

\unskip

%%%%%%%%%%%%%%%%%%%%%%%%%%%%%%%%%%%%%%%%%%%%%%%%%%%%%%%%%%%%%%%%%%%%%%%%%%%
\begin{table}[H]
\caption{Maximum efficiency of magnetic Penrose process and corresponding energy extraction regime for some typical radioactive decay modes in the vicinity of a black hole of mass $10~M_\odot$ with the spin parameter $a=0.8M$. Magnetic field is aligned along the rotation axis and has the strength $10^4$  G. Initial energy of decaying particle is taken to be of order of its rest mass, $m_X c^2$. In the case of the pair production, the energy of photon is taken to be $E_{\gamma} = 2 m_e c^2$ with recoil term neglected for simplicity.
\label{tab2}
}
%\begin{center}
\centering
\begin{tabular}{c  c c c c }
%\hline
\toprule
\textbf{Decay Mode }   &  \textbf{Generic Equation}  & \textbf{ Esc. p.} & \textbf{Efficiency} \boldmath$\eta_{\rm max}$ & \textbf{Regime of MPP}    	  \\	 										
%\hline%\hline 
\midrule

\multirow{6}{*}{$\alpha$ decay}	   &  \multirow{2}{*}{ $^{A}_{Z}$X$^0 \rightarrow \, ^{A-4}_{Z-2}$Y$^{2-} + \, ^{4}_{2}\alpha^{2+} $ } & Y  & <$0$ & - \\	
&  & $\alpha$ & $2824/A$ & ultra   \\	
&  \multirow{2}{*}{ $^{A}_{Z}$X$^+ \rightarrow \, ^{A-4}_{Z-2}$Y$^{-} + \, ^{4}_{2}\alpha^{2+} $ } & Y  &<$0$ & -- \\	
&  & $\alpha$ & $\sim$1 & moderate   \\
&  \multirow{2}{*}{ $^{A}_{Z}$X$^- \rightarrow \, ^{A-4}_{Z-2}$Y$^{3-} + \, ^{4}_{2}\alpha^{2+} $ } & Y  & $\sim$2 & moderate \\	
&  & $\alpha$ & $<0$ & -   \\
% \hline
\midrule

 \multirow{3}{*}{$\beta^-$ decay}	   &  \multirow{3}{*}{ $^{A}_{Z}$X$^0 \rightarrow \, ^{\,\,\,\,\,\,\,\,A}_{Z+1}$Y$^{+} + \, e^{-} + \bar{\nu}$ } & Y  & $1412/A$ & ultra \\	

 & & $e^-$ & <$0$ & --   \\				
 & & $\bar{\nu}$ & $ 0.06$ & low   \\				
% \hline
\midrule

\multirow{3}{*}{ $\beta^+$ decay}	   &  \multirow{3}{*}{ $^{A}_{Z}$X$^+ \rightarrow \, ^{\,\,\,\,\,\,\,\,A}_{Z-1}$Y$^{0} + \, e^+  + \nu $ } & Y  & $<$0 & -- \\	

 & & $e^+$ & $\sim$0 & low$/$--   \\	
 & & $\nu$ & <$0$ & --   \\	
% \hline
\midrule
 
 \multirow{2}{*}{$\gamma$ emission}	   &  \multirow{2}{*}{ $^{A}_{Z}$X$^0 \rightarrow ^{A}_{Z}$X'$^0 + \, ^{0}_{0}\gamma^{0} $ } & X'  & $0.06$ & low \\	

 & & $\gamma$ & $0.06$ & low   \\	
% \hline
\midrule

\multirow{2}{*}{Pair production}	   &  \multirow{2}{*}{ $\gamma^{0} \rightarrow e^- + e^+  $ } & { $e^-$ }  & { <$0$ } & { -- } \\
& & { $e^+$ }  & { $2.6 \times 10^6$ } & { ultra } \\
	
% \hline
\bottomrule

%\hline
\end{tabular}

%\end{center}
\end{table}
%%%%%%%%%%%%%%%%%%%%%%%%%%%%%%%%%%%%%%%%%%%%%%%%%%%%%%%%%%%%%%%%%%%%%%%%%%%

%%%%%%%%%% UHECR %%%%%%%%%%%%
\section{Astrophysical Applications \label{sec4}}
\unskip

\subsection{Ultra High Energy Cosmic Rays}
\unskip
\subsubsection{State of the Art}

One of the fascinating  applications of MPP is the explanation of the origin and production mechanism of ultra-high-energy cosmic rays (UHECR). These are extragalactic charged particles observed at energies above $10^{20}$~eV  \cite{2018ApJ...853L..29A,PAO:2017:sci:}. Detection of particles at such a high energy violates the GZK-cutoff limit \cite{Greisen:1966jv,Zatsepin:1966jv}, which requires exotic scenario for its production mechanism. 
There are indications that UHECRs are produced at AGNs evidenced by recent extrasolar neutrino detection by IceCube collaboration and simultaneous multi-wavelength observations in optical spectra tracing the source to blazar  \cite{2018Sci...361.1378I,2018Sci...361..147.}. 
The source TXS 0506+056 is located at the distance of $\sim$1.75  Gpc from Earth with relativistic jet directed towards us.
Recent observational data lean towards the mixed composition of UHECRs  \cite{2017arXiv170806592T,Aab:2016zth}, although all previous fluorescence measurements of cosmic rays demonstrated dominantly proton flux at high energies  \cite{2015APh....64...49A,2010PhRvL.104p1101A}.
 As observed by the largest cosmic ray telescopes, such as Pierre Auger Observatory  \cite{PAO:2017:sci:} and Telescope Array \cite{2017APh....86...21A}, the highest energy cosmic rays (>$10^{18}$ eV) are originated  outside of our Galaxy.
  Spectrum of cosmic rays demonstrate the presence of the knee at $\sim$$10^{15.5}$ eV and the ankle at $\sim$$10^{18.5}$ eV. Flux of particles above knee decreases considerably, which may indicate the change in the source from Galactic to extragalactic.
 The origin of cosmic rays above the knee and mechanism of production of highest energy cosmic rays remain under active debate. The~main difficulty in understanding of UHECR physics is extremely low flux of these particles. The~flux of UHECRs with energy $>$$10^{20}$eV is about one particle per km$^2$ per century. Therefore, recently, the new infrastructure for global search of cosmic rays called ’Cosmic Rays Extremely Distributed Observatory’ (CREDO) has been proposed, which is based on the detection of cosmic rays with large number of mobile smartphones around the globe \cite{CREDO:2018:}. 
 One of the promising acceleration scenarios of UHECRs is the shock acceleration in the plasma of relativistic jets (see, e.g.,~ \cite{2000PhST...85..191B}). However, due to extended acceleration length and interaction looses inside and outside of the jet, realization of this scenario in the nature is debatable. Below, we apply MPP for the explanation of UHECRs which in contrast to jet acceleration models does not require extended acceleration zone.   
 For this, we employ the beta-decay of free neutron in the vicinity of supermassive black hole immersed into external magnetic field. 
 { We   partially follow the results presented in  recent paper \cite{Tur-uhecr:2019:}. }

\subsubsection{Maximum Energy of Proton}

\begin{figure}[H]
  \centering
  \scalebox{0.9}[0.9]{
 \includegraphics[width=0.28\textwidth]{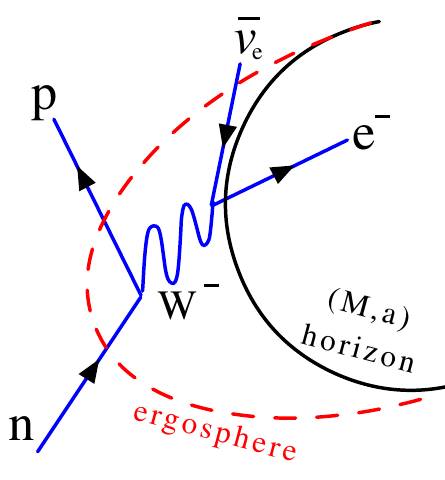}
 \includegraphics[width=0.36\textwidth]{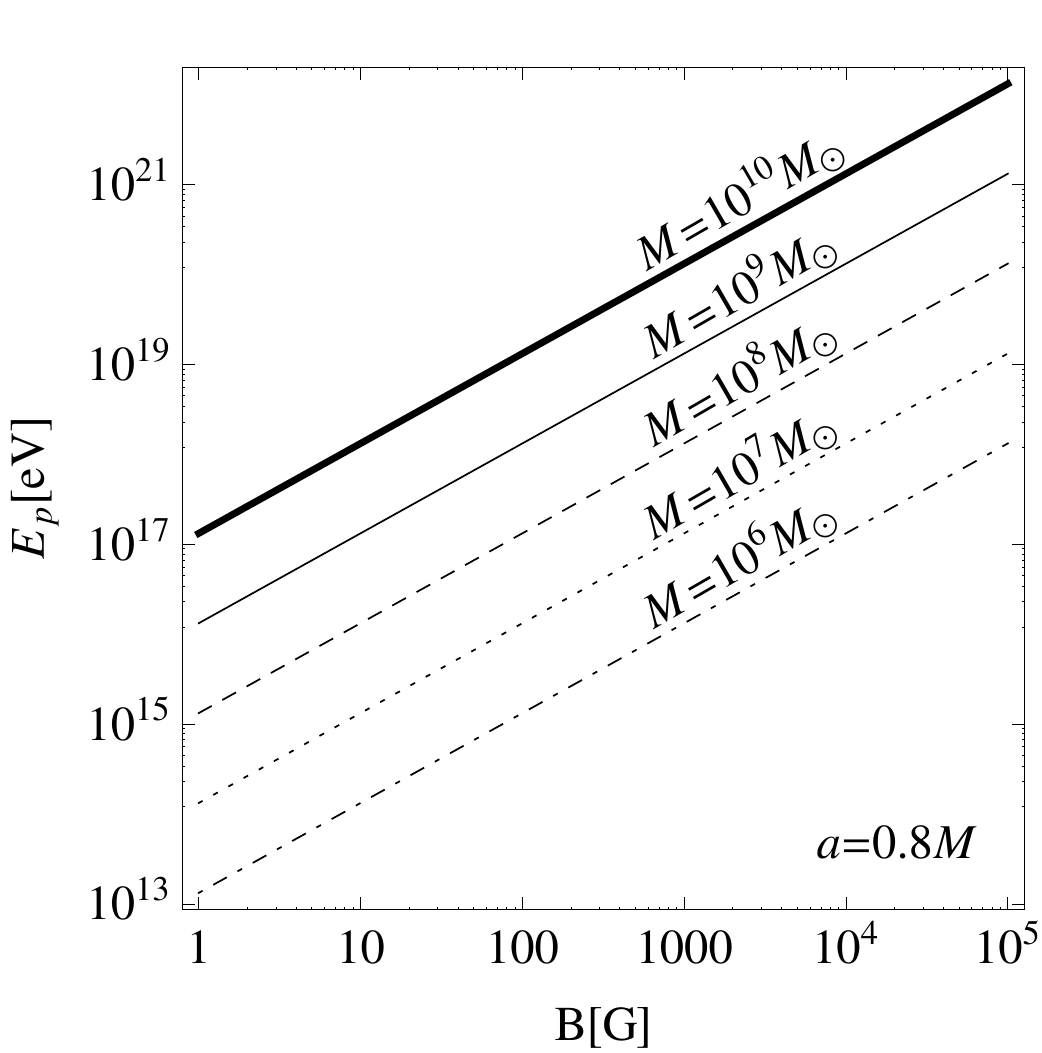}
 \includegraphics[width=0.36\textwidth]{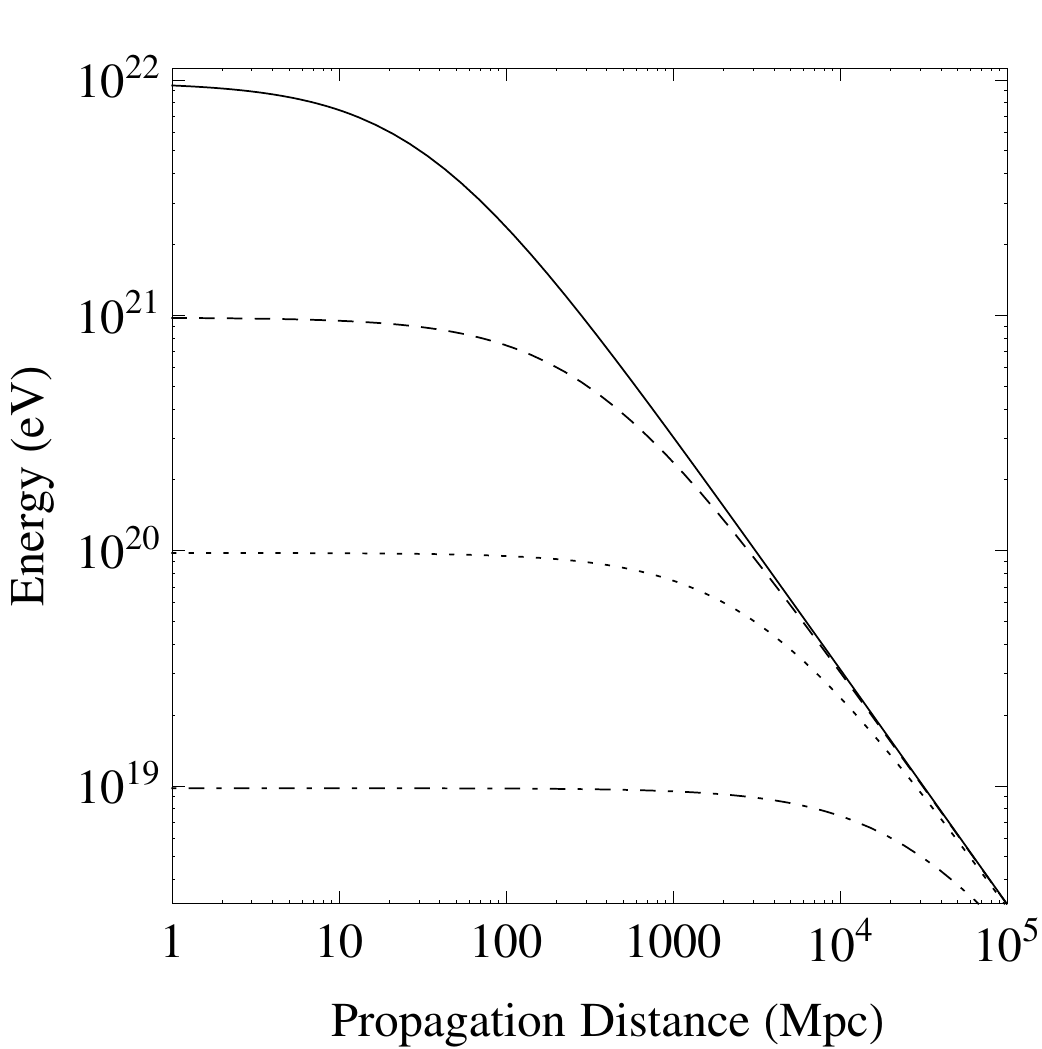}}
  \caption{(\textbf{Left})  Feynman diagram of the beta-decay in ergosphere of spinning black hole. Proton after decay is accelerated by induced black hole charge and escapes to infinity. (\textbf{Middle})  Proton energy in ultra-efficient MPP is plotted against  magnetic field for different values of black hole mass. (\textbf{Right}) The drop of proton energy over the distance comparable to local cosmological structures undergoing synchrotron radiation in a magnetic field of order of $10^{-5}$  G. }
  \label{fig_3}
\end{figure}

In Figure~\ref{fig_3} (left), we illustrate the Feynman diagram of the beta-decay in the ergosphere of rotating black hole which corresponds to MPP in ultra-efficient regime. Neutron decays into proton, electron and antineutrino. Due to expectation of alignment of accretion flow with a black hole rotation at least near the event horizon, gravitationally induced charge on the black hole is more plausibly positive. This implies that in the neutron beta-decay the escaping particle is proton. One should note that the effect of antineutrino on the energy of proton is negligible.

Applying MPP to the beta-decay of free neutron in vicinity of a standard  supermassive black hole (SMBH) with mass $M=10^9M_{\odot}$ and magnetic field of strength $B = 10^4$G, one can get proton escaping from black hole with energy
\begin{equation} \label{eq1}
%\centering
E_{ p^+} = 1.3 \times 10^{20} {\rm eV} \left( \frac{q}{e}\right) \left(\frac{m}{m_{p^+}}\right)^{-1} \left( \frac{B}{10^4 {\rm G}}\right) \left(\frac{M}{10^9 M_{\odot}} \right).
\end{equation}

We plot the dependence of proton's energy on magnetic field in Figure   \ref{fig_3} (middle) for various values of black hole masses. Spin is chosen to be $a=0.8M$ as an example. As  shown in Figure   \ref{fig-eff-ultra}, the spin of the black hole does not change the  efficiency of MPP dramatically, although higher the black hole spin, larger is the ergosphere, which enhances  probability of operation of MPP in realistic~conditions.

\subsubsection{Propagation of Cosmic Rays in Magnetic Field}

% GZK cut-off and synchrotron loses %
Depending on the constituents of primary cosmic rays, a significant amount of initial energy will be lost in the photo--pion interactions of cosmic ray particles with low energy photons of cosmic microwave background (CMB). Loss caused by such interactions will grow over cosmological distances if the energy of primary cosmic ray is larger than $\sim$$10^{19.7}$ eV. This effect of energy loss by primary cosmic rays is known as the GZK-cutoff \cite{Greisen:1966jv,Zatsepin:1966jv}. Observations of cosmic rays above GZK-cutoff limit indicate the location of sources within <$100$ Mpc. Indeed, the data obtained by Pierre Auger Observatory point to strong correlation between arrival direction of highest-energy cosmic rays with the location of nearby active galactic nuclei within a distance of $75$--$100$ Mpc from our Galaxy \cite{2007Sci...318..938P}. 

However, apart from CMB photons the UHECRs interact with magnetic fields along the propagation distance which can vary from intergalactic and Galactic magnetic fields of order $10^{-9}$--$10^{-5}$~G up to values exceeding $10^4$  G in the source region. Therefore it is important to investigate the synchrotron loss timescales and distances for UHECRs in various magnetic field. Cooling timescale of charged particle in curved spacetime is given by (\ref{life-time}) which has cubic dependence on mass of charged particle. Hence, electron cooling timescale is $\sim$$10^{10}$ shorter than protons, which makes heavier constituents more realistic cosmic ray candidates (see  Table~\ref{tab1}). In Figure   \ref{fig_3}~(right), we demonstrate energy loss of protons with various initial energies over propagation distance due to intereaction with Galactic order magnetic field $10^{-5}$  G. One can also see that in the case of neutron star (sometimes suggested as UHECRs source), whose magnetic field is typically above $10^{12}$  G, the timescale of synchrotron energy loss is extremely short even for protons and ions.

\subsubsection{CosmicRays from Galactic Center Black Hole}

Remarkably, MPP gives relatively precise prediction on the maximum proton energy produced by Galactic center black hole, which is the closest and best studied supermassive black hole candidate nowadays \cite{Eckart-etal:2017:FOP:}. The central black hole Sgr~A* has relatively accurate measurements of its mass being of $\sim$$4.1$ million solar masses based on the observations of surrounding stellar dynamics \cite{2017ApJ...845...22P}. Distance measurements are closely related to mass estimates which gives a distance of $\sim$$8.1$ kpc from Solar system to Galactic center. Estimate of spin is model dependent and requires observations of immediate regions of black hole. %
Based on multi-wavelength, radio, infrared and near-infrared interferometric data current constraints on the spin of Sgr~A* is  $a>0.4M$ \cite{2006A&A...455....1E}, although one cannot exclude near extremal values \cite{2018A&A...618L..10G}. Magnetic field around Sgr~A* has equipartition strength of $10$G \cite{Eckart-etal:2017:FOP:}, which can reach up to few hundred Gauss on the event horizon scale \cite{Eatough-etal:2013:Natur:}. 
Applying MPP to the beta-decay of neutron skirting close to Sgr~A*, we obtain the high-energy proton at the Galactic center with energy of~order
\beq
E_{p^+ \rm (SgrA^*)} \approx  10^{15.6} {\rm eV} \left( \frac{q}{e}\right) \left(\frac{m}{m_{p^+}}\right)^{-1} \left( \frac{B}{100 {\rm G}}\right) \left(\frac{a}{0.5 M} \right) \left(\frac{M}{4.14 \times 10^6 M_{\odot}} \right).
\eeq

Surprisingly, this energy has the same order of magnitude as the knee of the cosmic ray energy spectra, above which flux of particles demonstrate significant suppression \cite{2018ApJ...865...74A}.  Although many models have been suggested, the explanation of the origin of knee is still in strong debate due to number of uncertainties in Galactic magnetic field structure and lack of direct detections of primary cosmic rays. The sharpness of the knee energy spectra could in fact indicate the existence of a single source at knee energy level \cite{Erlykin_2005}. Thus, our model could serve as an alternative and relatively simple explanation of the knee appearing in the cosmic ray spectra.

The logical extension of the proposed model can be the search of correlations in the observational data of UHECRs with predictions of our model.

\subsection{Relativistic Jets}

Relativistic, collimated ejections of matter named jets have been observed in variety of objects, such as AGNs, X-ray  binaries, quasars,    etc. It is generally believed that the mechanism of production of jets is strongly connected with accretion processes in combined strong gravitational and electromagnetic fields. Being one of the fundamental problems of the modern relativistic astrophysics, the relativistic jets are under intensive considerations. Large interest is connected with attempts to understand how the energy from accreting matter is converted into the kinetic energy of escaping matter in a narrow cone, so that the jets acquire very large gamma-factors.
Here,   we describe a simple model of jet-like motion of particles due to the ionization of neutral Keplerian accretion disc in the equatorial plane of weakly magnetized rotating black hole. As we have already seen above, the ionization of neutral particles in the vicinity of black hole can provide ultra-high-energy to one of its charged fragments after the split, so the fragment has enough energy to escape from the black hole.   However, the motion of charged matter in the presence of magnetic field is always bounded in the equatorial plane (see, e.g., ~\cite{Tur-Stu-Kol:2016:PHYSR4:} and references therein). Therefore, the escape of particles from black hole is possible only along the magnetic field lines, which we assume to be aligned with the rotation axis of the black hole. This~assumption is well justified as any field configuration in the vicinity of black hole generally shares the symmetries of the background spacetime metric. Transition from equatorial motion of charged particles to the linear motion along the rotation axis occurs due to chaotic scattering of high-energy particles in the effective potential. This causes the interchange between the oscillatory and transnational modes of energy $E_{0} \rightarrow E_z$ of charged particles. The mathematical technique and numerical modelling of the chaotic scattering of charged particles by black hole is given in \cite{2016EPJC...76...32S}.

\subsubsection{Chaotic Scattering of Ionized Particles}

Motion of test particle is limited by the effective potential, so that $E = V_{\rm eff}$. While bounded in equatorial plane in the presence of magnetic field, the boundaries of motion can be open for escape to infinity along perpendicular direction to the equatorial plane.  A particle is able to escape to infinity if its energy is grater than  asymptotic value, which in terms of the specific energy $(\ce = E/m)$ is given by 
\beq
 \ce_{\rm min} = 
\Big\{ 
\begin{array}{l @{\quad} c @{\quad} l} 
2 a \cb + 1 & \textrm{for} & \cb \geq 0, \\ 
2 a \cb + \sqrt{1 - 4 \cb \cl} & \textrm{for} & \cb < 0, \\ 
\end{array} \label{minE}
\eeq 
where $\BB$ is defined by (\ref{cBB}). One can derive that 
\bea
 \ce_{\infty}^2 &=& \ce^2_{\rm z} + \ce^2_{\rm 0}, \label{ENflat1}  \\
 \ce^2_{\rm 0} &=& \dot{r}^2 + g_{\phi\phi} \omega^2 = \dot{r}^2 + \left( \cl/r - \cb r \right)^2 + 1, \label{ENflat2} \\
 \ce^2_{\rm z} &=& \dot{z}^2, \label{ENflat}
\eea
where $\ce_{\infty}$ is the energy of a particle measured at infinity. Thus, the total energy of a particle measured at infinity is composed from the longitudinal $\ce_{\rm z}$ and transverse $\ce_{\rm 0}$ parts.
It appears that near the black hole two components of energy $\ce_{\rm z}$ and $\ce_0$ are interchangeable \cite{2016EPJC...76...32S}, while in Minkowski spacetime the energies given by Equations (\ref{ENflat1})--(\ref{ENflat}) are integrals of motion and therefore cannot be transferred between two energy modes. In  Kerr spacetime metric the conserved quantity is the total covariant time component of momentum i.e., $\ce_{\infty}$. Interchange in two energy modes imply the change of the velocity of charged particles along the magnetic field line, which can increase to extremely large values due to combination of MPP with chaotic scattering effect.

%%%%%%%%%%%%%%%%%%%%%%%%%%%%%%%%%%%%%%%%%%%%%%%%%%%%%%%%%%%%%%%%%%%%%%%%%%%%%%%%%%%%%%%%%%%%%%%%%%%%%%%%%%%%%

\subsubsection{Escape Velocity}

%%%%%%%%%%%%%%%%%%%%%%%%%%%%%%%%%%%%%%%%%%%%%%%%%%%%%%%%%%%%%%%%%%%%%%%%%%%%%%%%%%%%%%%%%%%%%%%%%%%%%%%%%%%%%

Asymptotic value of Lorentz gamma factor is related to the energy at infinity in the following~form
\beq
 \gamma = u^{t} = \frac{\d t}{\d \tau} = \ce + \frac{q}{m} A_{t} = \ce - 2 a \cb = \ce_\infty. \label{gammaDEF}
\eeq

Escape velocity  $u^{\rm z} = dz/(d\tau)$ or $v^{\rm z} = dz/(dt)$ and the Lorentz gamma in $z$-direction $\gamma_{\rm z}$ can be expressed in terms of the relation in Equation (\ref{ENflat1}) as follows
\beq
u_{\rm z} = {\cal E}_{\rm z}, \, \quad v_{\rm z} = \frac{{\cal E}_{\rm z}}{{\cal E}_\infty}, \quad \gamma_{\rm z} = \frac{1}{\sqrt{1-v_{\rm z}^2}} =\frac{{\cal E}_\infty}{{\cal E}_0}. \label{speedZ}
\eeq

The Lorentz factor is maximal when $\ce_0$ is minimal, i.e.
\bea
 \BB>0 &:& \gamma_{\rm z(max)} = \frac{{\cal E}_\infty}{{\cal E}_{\rm 0(min)}} = \ce_\infty, \label{speedZmaxP} \\
 \BB<0 &:& \gamma_{\rm z(max)} = \frac{{\cal E}_\infty}{\sqrt{1-4\BB\cl}} . \label{speedZmaxM} 
\eea

In case of maximal acceleration with positive $\BB$, the orbital velocity of particle around black hole vanishes and all energy in equatorial plane  transfers into kinetic energy in $z$ direction, so that $u^{\phi} = 0$, for $\BB>0$. In case $\BB<0$, for maximal acceleration, the orbital velocity tends to the limit $u^{\phi} = 2 \cb \cl$.

\begin{figure}[H]
\includegraphics[width=\hsize]{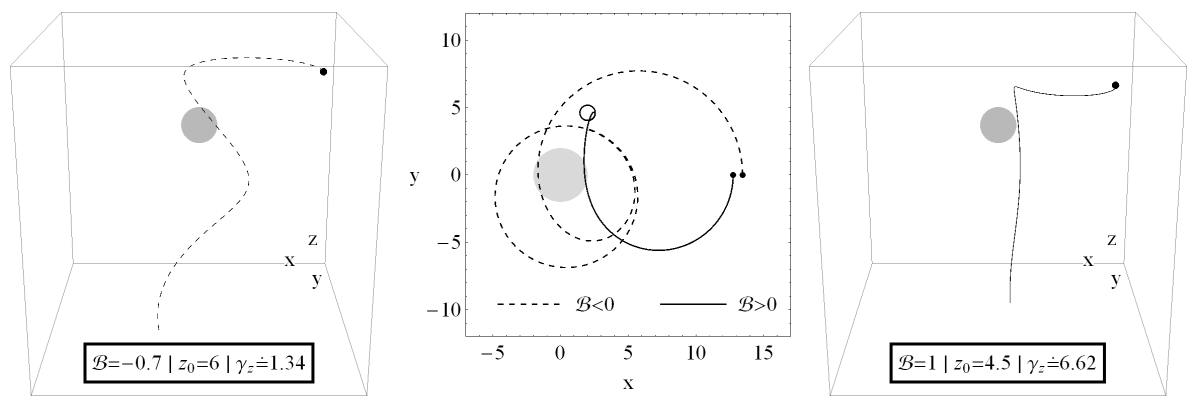}
\caption{ \label{largeGZ}
Trajectories of charged particles with negative (\textbf{left}) and positive (\textbf{right}) magnetic parameter $\BB = q B G M/(m c^4)$ escaping black hole along $z$ axis. Numerical values of corresponding Lorentz factors are given inside the plots. (\textbf{Middle}) The projections of both trajectories in the $x-y$ plane. The figure is reproduced from  \cite{2016EPJC...76...32S}.
}
\end{figure}

Comparison of trajectories in two cases with negative (left) and positive  (rights) magnetic parameter $\BB$ is represented in Figure   ~\ref{largeGZ}, which is reproduced from \cite{2016EPJC...76...32S}. The trajectories are found in  \cite{2016EPJC...76...32S} by numerical integration of the full set of equations of motion. Acceleration is larger in case of positive $\BB$.   Figure   ~\ref{largeGZ} represents the cases with relatively low magnetic parameters $|\BB| \sim 1$. For~elementary particles around astrophysical black holes, the magnetic parameter $\BB$ is usually very large (see, e.g.,    (\ref{estimation-BBsgra}) for electron around Sgr~A* black hole). Therefore, the Lorentz factor of escaping particles after decay following MPP can be extremely large. Energy of charged particles produced in MPP is mainly concentrated in $\ce_0$ mode due to considered formalism of decay in equatorial plane. The motion in that case is quasi-circular. Due to chaotic scattering all energy in $\ce_0$ mode can be transformed to longitudinal kinetic energy $\ce_{\rm z}$ in the black hole vicinity. Notable feature of the ``transmutation''  effect is that it does not require black hole's rotation and can operate also in Schwarzschild spacetime immersed into external magnetic field. However,     to produce high-energy particles the black hole rotation plays crucial role as it generates electric potential $A_t$ providing ultra-high acceleration to the charged particles.
 Thus, the combination of MPP with chaotic scattering effect for the acceleration of charged particles can serve as a simple model of relativistic jets as the model provides extremely large Lorentz factors ($\gamma_z \gg 1$) and strong collimation of escaping charged particles along the symmetry axis.

%-------------------------------------------------------------------------%
%-------------------------------------------------------------------------%

\subsection{Charge Separation in the Accretion Disk}

{ In the plasma setup, similar mechanism of energy extraction from rotating black holes can described by the mechanism of Ruffini and Wilson \cite{1975PhRvD..12.2959R}, based on the charge separation in a magnetized plasma of accretion disk around black hole. In this scenario, the black hole may act as a pulsar \cite{2018arXiv180807887L} with non-zero net charge of the black hole and resulting charged magnetosphere.} 
Plasma of accretion disk generally surrounding black holes is  usually considered to be neutral. This assumption is justified by the neutralization of any charged plasma configuration in relatively short timescales. However, taking an analogy of magnetosphere of pulsars, where the plasma acquires net charge density, called Goldreich--Julian (GJ) charge density \citep{1969ApJ...157..869G}, it is easy to show that similar situation occurs in the relativistically  moving plasma around black hole. In particular, given that the matter is ordinary two component fluid of hydrogen plasma, motion of plasma in relativistic speeds ${\bf v}$ in external magnetic field ${\bf B}$ induces an electric field component ${\bf D}$
\begin{equation} \label{eq-evB}
    {\bf D} = - {\bf v} \times {\bf B}.
\end{equation}

In the frame co-moving with plasma, electric current density is connected with electric field ${\bf D}$ by the Ohm's law, which takes the following form for an observer at rest
\begin{equation}
 {\bf j} = \gamma \sigma  \left( {\bf D} + {\bf v} \times {\bf B} - {\bf v} \cdot ({\bf v} \cdot {\bf D}) \right) + \rho {\bf v}.
\end{equation}

Since electric field in co-moving frame must be neutralized, which usually happens in a very short timescales, electric field with respect to observer at rest does not vanish. 
Velocity of an arbitrary point of the disk is ${\bf v} = {\bf \Omega} \times {\bf R}$, where $\Omega$ is angular velocity and $R$ is distance to chosen point of the disk. This implies that net charge density in disk is given by 
\begin{equation}
    \rho = - \frac{1}{4 \pi} \Omega B.
\end{equation}

Consequently, charge separation in  plasma moving around weakly magnetized rotating black hole can be applied to MPP in a similar way as decay of neutral particles into charged fragments, described in previous sections. General relativistic magnetohydrodynamic  simulations (GRMHD) of two component plasma fluid with global charge neutrality revolving around rotating black hole in external magnetic field should be tested as it may lead to  energy extraction from black hole in ultra efficient way.

%%%%%%%%%%%%%%%%%%%%%%%%%%%%%%%%%%%%%%%%%%
\section{Discussion} \label{sec-discussion}

High energy phenomena such as quasars, AGNs, GRBs, FRBs, UHECRs and so on ask for two things: (1) a very  large reservoir of energy; (2) that it be harnessed very efficiently to give enormous power output. A rotating black hole has $29\%$ its mass in rotational energy, which could be mined out. It therefore becomes the natural candidate for the huge reservoir we are looking for. Now,  the question is of very efficiently tapping   this rotational energy. 

In 1969, Roger Penrose proposed a very novel and  purely geometric process  \cite{Penrose:1969:NCRS:} of energy extraction from a rotating black hole. The key property driving the process was the existence of negative energy orbits relative to an observer at infinity in the vicinity of black hole horizon, the region called the ergosphere. At the root of all this is the property of frame-dragging by which black hole shares its rotation with surrounding space. That is, space around a rotating black hole has inherent rotation so as to give angular velocity even to particle having zero angular momentum. The crucial point is that this property of  existence of negative energy orbits in ergosphere is at the root of all processes involved in extracting energy from black hole. 

The question then boils down to finding the most efficient process. In the original Penrose process, the maximum possible efficiency was $20.7\%$. More important was the question that energy required to push a particle on NEOs is non-trivial---
relative velocity between fragments to be greater $1/2 c$~\cite{Bar-Pre-Teu:1972:APJ:,Wald:1974:APJ:}. This requirement was astrophysically insurmountable because there can be no astrophysical mechanism that could accelerate particles to such relativistic velocity almost instantaneously. Thus, Penrose process was very ingenious and interesting  but it was astrophysically not viable to power high energy sources. Then,   various variants of the process were considered \cite{Pir-Sha-Kat:1975:APJL:,Pir-Sha:1977:APJ:,Zaslavskii:2016:PRD:,Ber-Bri-Car:2015:PRL:,Schnittman2018,2016PhRvD..93d3015L,2009PhRvL.103k1102B,Fro:2012:PHYSR4:,2015PhRvD..91h4044P} but not of   any consequence or avail.  

In 1977, came the BZ \cite{Bla-Zna:1977:MNRAS:}, which is set with black hole sitting in a magnetic field whose field lines are twisted due to frame dragging producing quadrupolar  electric potential difference between pole and equator. Its discharge   drives out energy and angular momentum from the black hole. It is an astrophysically very exciting and interesting process. However,   NEOs also play the critical role in working of the process. Further, the process requires polarization of vacuum in ergosphere, which would require threshold magnetic field of order $10^4$ G. Here,   rotational energy of black hole is extracted electromagnetically---the magnetic field plays the role of a catalytic agent. 

In 1985, an electromagnetic version of Penrose process (MPP)   \cite{Wag-Dhu-Dad:1985:APJ:,Bha-Dhu-Dad:1985:JAPA:,Par-etal:1986:APJ:,Wagh-Dadhich:1989:PR:}   was considered and it was shown that the limit on threshold relative velocity could be easily overcome by particle's interaction with electromagnetic field; i.e., energy required for particle to ride on NEO could now come from $-qA_t$ leaving particle's velocity completely free. This was a wonderful revival of the process for astrophysical applications. It was only  shown  that the process is highly efficient, but so  much so that efficiency could exceed $100\%$ for as low field as milliGauss. This was for the first demonstration of efficiency exceeding $100\%$.  

Clearly, BZ and MPP share the same setting of black hole immersed in magnetic field,  so are the two the same? True, in moderately high magnetic fild range, the two seem to be similar. Recently, it has been argued \cite{2018MNRAS.478L..89D} that MPP is a general process  that  works in all magnetic field range while BZ requires the threshold magnetic field, and for that range the former approximates to the latter. That is BZ is contained in MPP in high field range. 

All this was of course in the linear test particle accretion regime, which is admittedly an idealized situation. More realistic setting could not be taken up due to lack of computing power at that time. Since around 2010, fully relativistic magnetohydrodynamic flow simulation models have been studied~\cite{Nar-McC-Tch:book:2014:,Tch-Nar-McK:2011a:MNRAS:}; it is remarkable that the process continues to remain highly  efficient, efficiency ranging to $300\%$. It turns out that MPP/BZ is the most plausible and important powering mechanism for high energy sources such as quasars,    etc. It is remarkable and very gratifying for one of us that the mid-1980s prediction of efficiency exceeding $100\%$ has been beautifully borne out. Thus, MPP is perhaps the key process of mining rotational energy of black hole and powering the high energy~sources. 

Among astrophysical phenomena which can be directly related to MPP are the  ultra-high-energy cosmic rays and relativistic jets. In the case of UHECRs, MPP is able to produce protons with energy exceeding  $10^{20}$ eV employing neutron beta-decay in the ergosphere of supermassive black hole of mass $10^9 M_{\odot}$ and magnetic field of $10^4$  G. Heavier cosmic ray constituents of similar energy range can  also be achieved depending on the decay mode of infalling neutral matter. We summarize  some of the potential radioactive decay modes with ultra efficient energy extraction in Table~\ref{tab2}.  
 Depending on the energy range and constituents of primary UHECRs, this can provide constraints on the black hole's mass, magnetic field and the distance to the source candidate. Remarkably, the knee energy of the cosmic ray spectrum at $\sim$$10^{15.5}$ eV coincides with the maximum proton energy obtained by MPP applied to the Galactic center black hole Sgr~A*.  Correlations of detected cosmic rays with nearby SMBHs of mass and magnetic field predicted by MPP needs to be explored. 
 
 Similarly, MPP can provide  jet-like behavior of charged particles, namely the large Lorentz gamma factor and strong collimation of ejected matter. This can be related to the ionization of accretion disk either by decay of infalling neutral particle into charged fragments or due to charge separation in a relativistic plasma of accretion disk. In the latter scenario black hole may behave as a pulsar \cite{2018arXiv180807887L} with  net stable charge of both black hole and surrounding magnetosphere with respect to observer at rest at infinity. Ejection of ionized particles to infinity takes place along magnetic field lines, which can be open to infinity at the polar caps of black holes. 
  Then, charged particles escape to infinity due to chaotic scattering effect caused by the interchange between the oscillatory and translational modes of the total energy of escaping particles $E_0 \rightarrow E_z$  \cite{2016EPJC...76...32S}. Thus, the combination of MPP with chaotic scattering effect leads to the high Lorentz factors and alignment of trajectories of escaping particles along the rotation axis of a black hole. This is quite promising and exciting that MPP may find further astrophysical applications in the near future, such as to be relevant for the operation of many other high-energy astrophysical phenomena, such as  GRBs and FRBs, among others.

Finally, we have come long way   from 1969---a good half a century---and it is remarkable to see what was proposed as a thought experiment has come to stay firmly and beautifully as one of the prime movers in  high energy astrophysical phenomena. It is an excellent example of purely geometry driven process aided by magnetic field could wonderfully serve as powering engine for such complex astrophysical systems such as quasars, AGNs, UHECR, and so on. In this review, we have attempted to recount the historical evolution of the process, and it is fair to say that it has more borne out the promise it held.

%%%%%%%%%%%%%%%%%%%%%%%%%%%%%%%%%%%%%%%%%%
%\section{Conclusions}

%This section is not mandatory, but can be added to the manuscript if the discussion is unusually long or complex.

%%%%%%%%%%%%%%%%%%%%%%%%%%%%%%%%%%%%%%%%%%
\vspace{6pt} 

%%%%%%%%%%%%%%%%%%%%%%%%%%%%%%%%%%%%%%%%%%
%% optional
%\supplementary{The following are available online at \linksupplementary{s1}, Figure S1: title, Table S1: title, Video S1: title.}


%%%%%%%%%%%%%%%%%%%%%%%%%%%%%%%%%%%%%%%%%%
\authorcontributions{Both authors have contributed equally to the work.}

%%%%%%%%%%%%%%%%%%%%%%%%%%%%%%%%%%%%%%%%%%
\funding{A.T. was supported by International Mobility Project CZ.02.2.69/0.0/0.0/16\_027/0008521.
%the Czech Science Foundation Grant No. 16-03564Y and internal grant of the Silesian University SGS/14/2016.
}

%%%%%%%%%%%%%%%%%%%%%%%%%%%%%%%%%%%%%%%%%%
\acknowledgments{The authors would like to acknowledge the Institutional support of
the Insitute of Physics of the Silesian University in Opava. We also thank Zden{\v e}k Stuchl\'ik, Martin Kolo{\v s}, Bobomurat Ahmedov, Andreas Eckart for useful discussions.}

%%%%%%%%%%%%%%%%%%%%%%%%%%%%%%%%%%%%%%%%%%
\conflictsofinterest{ The authors declare no conflict of interest.} 

%%%%%%%%%%%%%%%%%%%%%%%%%%%%%%%%%%%%%%%%%%
%% optional
\abbreviations{The following abbreviations are used in this manuscript:\\

\noindent 
\begin{tabular}{@{}ll}
MPP & magnetic Penrose process\\
BZ & Blandford--Znajek mechanism\\
SMBH & supermassive black hole\\
UHECR & ultra-high-energy cosmic ray\\
MF & magnetic field\\
NEO & negative energy orbit\\
AGN & active galactic nuclei\\
FRB & fast radio burst\\
GRB & gamma ray burst
\end{tabular}}

\reftitle{References}

%%%%%%%%%%%%%%%%%%%%%%%%%%%%%%%%%%%%%%%%%%
%% optional
%\sampleavailability{Samples of the compounds ...... are available from the authors.}

%% for journal Sci
%\reviewreports{\\
%Reviewer 1 comments and authors’ response\\
%Reviewer 2 comments and authors’ response\\
%Reviewer 3 comments and authors’ response
%}

%%%%%%%%%%%%%%%%%%%%%%%%%%%%%%%%%%%%%%%%%%
\end{document}